\documentclass{egpubl}
\usepackage{egsr2025}
 
\SpecialIssuePaper         

\CGFccby

\usepackage[utf8]{inputenc}									
\usepackage[T1]{fontenc}									
\usepackage{dfadobe}

\usepackage{cite}  
\BibtexOrBiblatex
\electronicVersion
\PrintedOrElectronic

\ifpdf \usepackage[pdftex]{graphicx} \pdfcompresslevel=9
\else \usepackage[dvips]{graphicx} \fi

\usepackage{egweblnk}

\newif\ifanonymous
\anonymousfalse

\title[Real-time IBL Glints]%
      {Real-Time Image-based Lighting of Glints}

\ifanonymous
\author[paper1034]
{\parbox{\textwidth}{\centering
		paper1034
	} \\
	{\parbox{\textwidth}{\centering }}
}
\else
\author[Tom Kneiphof \& Reinhard Klein]
{\parbox{\textwidth}{\centering
		Tom Kneiphof$^{1}$\orcid{0000-0002-7237-2431}
		and Reinhard Klein$^{1}$\orcid{0000-0002-5505-9347}
	} \\
	{\parbox{\textwidth}{\centering $^1$University of Bonn, Germany}}
}
\fi

\usepackage{amsmath}    
\usepackage{amssymb}    
\usepackage{mathtools}
\usepackage{bm}         
\usepackage{cancel}     
\usepackage{nicefrac}   

\usepackage{siunitx}    
\usepackage{csquotes}

\usepackage{subcaption}
\usepackage{algorithm}
\usepackage{algpseudocode}

\usepackage{tikz}
\usepackage{tikz-3dplot}
\usepackage{tikz-3dplot-circleofsphere}
\usetikzlibrary{positioning,calc}
\usetikzlibrary{shapes.misc}
\usetikzlibrary{arrows.meta}
\usetikzlibrary{patterns}
\usetikzlibrary{decorations.pathreplacing,calligraphy}
\tikzset{cross/.style={cross out, draw=black, minimum size=2*(#1-\pgflinewidth), inner sep=0pt, outer sep=0pt},cross/.default={1pt}}
\pgfdeclarelayer{bg}    
\pgfsetlayers{bg,main}  

\usepackage{pgfplots}
\usepackage{pgfplotstable}
\usepgfplotslibrary{polar}
\usepgfplotslibrary{fillbetween}

\usepackage[l3]{csvsimple}

\usepackage[toc]{appendix}

\usepackage{booktabs}

\usepackage{url}
\usepackage{cleveref}
\crefname{equation}{Eq.}{Eqs.}
\crefname{figure}{Fig.}{Figs.}
\crefname{table}{Tab.}{Tabs.}
\crefname{section}{Sec.}{Secs.}
\crefname{algorithm}{Alg.}{Algs.}
\crefname{appendix}{Appendix}{Appendices}

\newcommand{\hide}[1]{}

\definecolor{revisioncolor}{RGB}{34,139,34}
\ifanonymous
\newcommand{\revised}[1]{\textcolor{revisioncolor}{{}\ifx&#1&[\ldots]\else#1\fi{}}}
\newenvironment{revisedblock}
	{\color{revisioncolor}} 
	{} 
\else
\newcommand{\revised}[1]{#1}
	{} 
\fi

\newif\iffigsinline
\newif\iffigsatend
\figsinlinetrue
\figsatendfalse

\newcommand{\citeq}[1]{\citeleft\citen{#1}\citeright}

\let\originalleft\left
\let\originalright\right
\renewcommand{\left}{\mathopen{}\mathclose\bgroup\originalleft}
\renewcommand{\right}{\aftergroup\egroup\originalright}

\newcommand{\tikzboxed}[2][]{\begin{tikzpicture}[baseline=(current bounding box.base)]\node[anchor=base,draw,#1]{$\displaystyle #2$};\end{tikzpicture}}

\DeclareMathOperator{\clamp}{clamp}
\DeclareMathOperator{\lerp}{lerp}

\newcommand{\dx}[1]{\;\mathrm{d}#1}
\newcommand{\sprod}[2]{\langle #1 \,, #2 \rangle}
\newcommand{\derivative}[2]{{ \frac{\partial#1}{\partial#2} }}

\newcommand{\Ex}[2][]{\mathbb{E}_{#1}\left[#2\right]}

\newcommand{\srof}[1]{{|#1|}}

\newcommand{\dirH}{{\mathbf{h}}}
\newcommand{\dirM}{{\mathbf{m}}}
\newcommand{\dirN}{{\mathbf{n}}}

\newcommand{\dirI}{{\omega_i}}
\newcommand{\dirO}{{\omega_o}}
\newcommand{\dirL}{{\omega_i}}

\newcommand{\dirV}{{\omega_o}}
\newcommand{\dirR}{{\omega_r}}
\newcommand{\pos}{{\mathbf{x}}}

\newcommand{\dirHk}[1][k]{{\dirH^{(#1)}}}
\newcommand{\posk}[1][k]{{\pos^{(#1)}}}

\newcommand{\domainIsup}[1][]{{\ifx&#1&\Omega_i\else\Omega_i^{#1}\fi}}
\newcommand{\domainI}{{\domainIsup}}
\newcommand{\domainHsup}[1][]{{\ifx&#1&\Omega_h\else\Omega_h^{#1}\fi}}
\newcommand{\domainH}{{\domainHsup}}

\newcommand{\domainHemi}{{\mathcal{H}}}
\newcommand{\domainSph}{{S^2}}

\newcommand{\domainIk}[1][k]{\domainIsup[(#1)]}
\newcommand{\domainHk}[1][k]{\domainHsup[(#1)]}

\newcommand{\hdotl}{{ \sprod{\dirH}{\dirL} }}

\newcommand{\ndotl}{{ \sprod{\dirN}{\dirL} }}
\newcommand{\ndotv}{{ \sprod{\dirN}{\dirV} }}

\newcommand{\pxFoot}{{\mathcal{P}}}

\newcommand{\brdfSymb}[1][]{{#1\rho}}
\newcommand{\brdfArgs}[2][]{{\brdfSymb[#1]\left(#2\right)}}
\newcommand{\brdf}[1][]{{\brdfArgs[#1]{\dirI,\dirO}}}

\newcommand{\brdfnolSymb}[1][]{{\brdfSymb[\tilde]}}
\newcommand{\brdfnolArgs}[2][]{{\brdfnolSymb[#1]\left(#2\right)}}
\newcommand{\brdfnol}[1][]{{\brdfnolArgs[#1]{\dirI,\dirO}}}

\newcommand{\fresnelSymb}[1][]{{#1F}}
\newcommand{\fresnelArgs}[2][]{{\fresnelSymb[#1]\left(#2\right)}}
\newcommand{\fresnel}[1][]{{\fresnelArgs[#1]{\dirI, \dirH}}}

\newcommand{\geomSymb}[1][]{{#1G}}
\newcommand{\geomArgs}[2][]{{\geomSymb[#1]\left(#2\right)}}
\newcommand{\geom}[1][]{{\geomArgs[#1]{\dirI, \dirO}}}

\newcommand{\ndfSymb}[1][]{{#1D}}
\newcommand{\ndfArgs}[2][]{{\ndfSymb[#1]\left(#2\right)}}
\newcommand{\ndf}[1][]{{\mathop{\ndfArgs[#1]{\dirH}}}}

\newcommand{\ndfFootSymb}{{\ndfSymb[\hat]_\pxFoot}}
\newcommand{\ndfFootArgs}[1]{{\ndfFootSymb\left(#1\right)}}
\newcommand{\ndfFoot}{{\ndfFootArgs{\dirH}}}

\newcommand{\ndfholSymb}[1][]{{\ndfSymb[\tilde]}}
\newcommand{\ndfholArgs}[2][]{{\ndfholSymb[#1]\left(#2\right)}}
\newcommand{\ndfhol}[1][]{{\ndfholArgs[#1]{\dirH}}}

\newcommand{\ndfholFootSymb}{{\ndfSymb[\tilde]_\pxFoot}}
\newcommand{\ndfholFootArgs}[1]{{\ndfholFootSymb\left(#1\right)}}
\newcommand{\ndfholFoot}{{\ndfholFootArgs{\dirH}}}

\newcommand{\LiSymb}[1][]{{{#1L}_i}}
\newcommand{\LiArgs}[2][]{{\LiSymb[#1]\left(#2\right)}}
\newcommand{\Li}[1][]{{\LiArgs[#1]{\dirI}}}

\newcommand{\totalLi}{{\LiSymb[]}}

\newcommand{\Lik}[1][k]{{L_i^{(#1)} }}
\newcommand{\wkSymb}[1][k]{{w^{(#1)} }}
\newcommand{\wkArgs}[2][k]{{\wkSymb[#1]\left(#2\right) }}
\newcommand{\wk}[1][k]{{\wkArgs[#1]{\dirI} }}

\newcommand{\LoSymb}[1][]{{{#1L}_o}}
\newcommand{\LoArgs}[2][]{{\LoSymb[#1]\left(#2\right)}}
\newcommand{\Lo}[1][]{{\LoArgs[#1]{\dirO}}}

\newcommand{\LoFootSymb}{{\hat{L}_{o,\pxFoot}}}
\newcommand{\LoFootArgs}[1]{{\LoFootSymb\left(#1\right)}}
\newcommand{\LoFoot}{{\LoFootArgs{\dirO}}}

\newcommand{\Nfoot}{{N_\pxFoot}}
\newcommand{\Ntotal}{{N_0}}
\newcommand{\Ndensity}{{\rho_N}}
\newcommand{\probHsup}[1][]{{\ifx&#1&p_{\domainH}\else p_{\domainH}^{#1}\fi}}
\newcommand{\probH}{{\probHsup}}
\newcommand{\probHk}[1][k]{{\probHsup[(#1)]}}
\newcommand{\probHvec}[1][]{{\mathbf{p}_{\domainH}}}
\newcommand{\intDLong}[1][]{{ \int_{\domainHsup[#1]} \ndf \dx{\dirH} }}
\newcommand{\intDHemiLong}{{ \int_\domainHemi \ndf \dx{\dirH} }}
\newcommand{\intD}[1][]{{ \mathcal{D}_{\domainHsup[#1]} }}
\newcommand{\intDk}[1][]{{ \mathcal{D}^{(k)}_{\domainHsup[#1]} }}
\newcommand{\intDHemi}{{ \mathcal{D}_\domainHemi }}

\newcommand{\brdfFGD}{{E_{\brdfSymb}\left(\dirO\right)}}
\newcommand{\ndfFGD}{{E_{\ndfSymb}\left(\dirO\right)}}

\newcommand{\meanRefl}{{\mathcal{FG}}}

\newcommand{\iblKernel}{{p_{\LiSymb}(\dirI)}}

\newcommand{\npos}{{n}}
\newcommand{\nneg}{{\bar{n}}}

\newcommand{\Panyhit}{{P_{\npos\geq1}}}
\newcommand{\Pallhit}{{P_{\npos=N}}}
\newcommand{\Ponehit}{{P_{\npos=1}}}
\newcommand{\Pgauss}{{P_{G}}}

\newcommand{\Pallhitneg}{{P_{\nneg=N}}}
\newcommand{\Ponehitneg}{{P_{\nneg=1}}}

\newcommand{\maskSymb}{{B}}
\newcommand{\maskallhit}{{\maskSymb_{\npos=N}}}

\newcommand{\maskalloronehit}{{\maskSymb_{\npos\in\{1,N\}}}}
\newcommand{\maskallhitneg}{{\maskSymb_{\nneg=N}}}

\newcommand{\maskalloronehitneg}{{\maskSymb_{\nneg\in\{1,N\}}}}
\newcommand{\maskgauss}{{\maskSymb_{G}}}

\newcommand{\Ngeqone}{{N_{\geq1}}}
\newcommand{\Ngeqtwo}{{N_{\geq2}}}

\definecolor{myred}{RGB}{228,26,28}

\definecolor{set1red}{RGB}{228,26,28}
\definecolor{set1green}{RGB}{77,175,74}
\definecolor{set1blue}{RGB}{55,126,184}
\definecolor{set1violet}{RGB}{152,78,163}
\definecolor{set1orange}{RGB}{255,127,0}

\definecolor{set2green}{RGB}{102,194,165}
\definecolor{set2orange}{RGB}{252,141,98}
\definecolor{set2blue}{RGB}{141,160,203}
\definecolor{set2violet}{RGB}{231,138,195}
\definecolor{set2lightgreen}{RGB}{166,216,84}
\definecolor{set2yellow}{RGB}{255,217,47}
\definecolor{set2beige}{RGB}{229,196,148}
\definecolor{set2grey}{RGB}{179,179,179}

\definecolor{pastel2green}{HTML}{b3e2cd}
\definecolor{pastel2red}{HTML}{fdcdac}
\definecolor{pastel2blue}{HTML}{cbd5e8}

\newcommand{\mytdplotsetcoordclipped}[4]{%
	\tdplotsetcoord{#1}{#2}{#3}{#4}%
	\pgfmathsetmacro{\tdplotresx}{\raarot * \stcpv + \rabrot * \stspv + \racrot * \costhetavec}%
	\pgfmathsetmacro{\tdplotresy}{\rbarot * \stcpv + \rbbrot * \stspv + \rbcrot * \costhetavec}%
	\pgfmathsetmacro{\tdplotresz}{\rcarot * \stcpv + \rcbrot * \stspv + \rccrot * \costhetavec}%
	\pgfmathparse{\tdplotresz >= 0}%
	\ifthenelse{\equal{\pgfmathresult}{1}}{%
		\pgfmathsetmacro{\rscreen}{1}%
	}{%
		\pgfmathsetmacro{\rscreen}{sqrt(\tdplotresx*\tdplotresx+\tdplotresy*\tdplotresy)}%
	}%
	\coordinate (#1boundary) at ($1/\rscreen*(#1)$);

}
\newcommand{\mytdplotsetcoordclippedext}[5][1.5]{%
	\mytdplotsetcoordclipped{#2}{#3}{#4}{#5}
	\tdplotsetcoord{#2ext}{#1*#3}{#4}{#5}
}

\newcommand{\mytdplotdrawhemisphereoutline}[2][]{%
	\pgfmathsetmacro{\rx}{#2}
	\pgfmathsetmacro{\ry}{\rx*cos(\tdplotmaintheta)}
	\path[tdplot_screen_coords,#1] (0:\rx) arc (0:180:\rx) arc (180:360:{\rx} and \ry) -- cycle;
}
\newcommand{\mytdplotdrawhemisphereclip}[2][]{%
	\pgfmathsetmacro{\rx}{#2}
	\pgfmathsetmacro{\ry}{\rx*cos(\tdplotmaintheta)}
	\clip[tdplot_screen_coords,#1] (0:\rx) arc (0:180:\rx) arc (180:360:{\rx} and \ry) -- cycle;
}
\newcommand{\mytdplotdrawhemispherebg}[2]{%
	\begin{scope}
		\mytdplotdrawhemisphereclip{#1}
		\pgfmathsetmacro{\mycy}{#1*(1-cos(\tdplotmaintheta))/2}
		\pgfmathsetmacro{\myw}{2*#1}
		\pgfmathsetmacro{\myh}{#1*(1+cos(\tdplotmaintheta))}
		\node[tdplot_screen_coords] at (0, \mycy, 0) {\includegraphics[width=\myw cm,height=\myh cm]{#2}};
	\end{scope}
}

\newcommand{\mytdplotdrawhemisphereoutlinerot}[4][]{%
	\tdplotsinandcos{\sinthetavec}{\costhetavec}{#3}%
	\tdplotsinandcos{\sinphivec}{\cosphivec}{#4}%
	\tdplotmult{\stcpv}{\sinthetavec}{\cosphivec}%
	\tdplotmult{\stspv}{\sinthetavec}{\sinphivec}%
	\tdplottransformmainscreen{\stcpv}{\stspv}{\costhetavec}
	\pgfmathsetmacro{\nx}{\tdplotresx}
	\pgfmathsetmacro{\ny}{\tdplotresy}
	\pgfmathsetmacro{\nangle}{atan2(-\nx, \ny)}
	\pgfmathsetmacro{\nsqnrm}{\nx*\nx + \ny*\ny}
	\pgfmathsetmacro{\rx}{#2}
	\pgfmathsetmacro{\ry}{\rx*sqrt(1-\nsqnrm)}
	\path[tdplot_screen_coords,rotate=\nangle,#1] (0:\rx) arc (0:180:\rx) arc (180:360:{\rx} and \ry) -- cycle;
}
\newcommand{\mytdplotdrawhemispherecliprot}[4][]{%
	\tdplotsinandcos{\sinthetavec}{\costhetavec}{#3}%
	\tdplotsinandcos{\sinphivec}{\cosphivec}{#4}%
	\tdplotmult{\stcpv}{\sinthetavec}{\cosphivec}%
	\tdplotmult{\stspv}{\sinthetavec}{\sinphivec}%
	\tdplottransformmainscreen{\stcpv}{\stspv}{\costhetavec}
	\pgfmathsetmacro{\nx}{\tdplotresx}
	\pgfmathsetmacro{\ny}{\tdplotresy}
	\pgfmathsetmacro{\nangle}{atan2(-\nx, \ny)}
	\pgfmathsetmacro{\nsqnrm}{\nx*\nx + \ny*\ny}
	\pgfmathsetmacro{\rx}{#2}
	\pgfmathsetmacro{\ry}{\rx*sqrt(1-\nsqnrm)}
	\clip[tdplot_screen_coords,rotate=\nangle,#1] (0:\rx) arc (0:180:\rx) arc (180:360:{\rx} and \ry) -- cycle;
}
\newcommand{\mytdplotdrawhemispherebgrot}[4]{%
\begin{scope}
	\mytdplotdrawhemispherecliprot{#1}{#2}{#3}
	\tdplotsinandcos{\sinthetavec}{\costhetavec}{#2}%
	\tdplotsinandcos{\sinphivec}{\cosphivec}{#3}%
	\tdplotmult{\stcpv}{\sinthetavec}{\cosphivec}%
	\tdplotmult{\stspv}{\sinthetavec}{\sinphivec}%
	\tdplottransformmainscreen{\stcpv}{\stspv}{\costhetavec}
	\pgfmathsetmacro{\nx}{\tdplotresx}
	\pgfmathsetmacro{\ny}{\tdplotresy}
	\pgfmathsetmacro{\nsqnrm}{\nx*\nx + \ny*\ny}
	\pgfmathsetmacro{\tx}{-\ny/sqrt(\nsqnrm)}
	\pgfmathsetmacro{\ty}{\nx/sqrt(\nsqnrm)}
	\pgfmathsetmacro{\bx}{\nx*sqrt((1-\nsqnrm)/\nsqnrm)}
	\pgfmathsetmacro{\by}{\ny*sqrt((1-\nsqnrm)/\nsqnrm)}
	\pgfmathsetmacro{\ewaC}{\tx*\tx+\bx*\bx}
	\pgfmathsetmacro{\ewaA}{\ty*\ty+\by*\by}
	\pgfmathsetmacro{\ewaB}{-2*(\tx*\ty+\bx*\by)}
	\pgfmathsetmacro{\ewaF}{\ewaC*\ewaA - \ewaB*\ewaB*0.25}
	\pgfmathsetmacro{\ewaA}{\ewaA/\ewaF}
	\pgfmathsetmacro{\ewaB}{\ewaB/\ewaF}
	\pgfmathsetmacro{\ewaC}{\ewaC/\ewaF}
	\pgfmathsetmacro{\ewaDet}{-\ewaB*\ewaB+4*\ewaC*\ewaA}
	\pgfmathsetmacro{\ewaX}{2*sqrt(\ewaDet*\ewaC)/\ewaDet}
	\pgfmathsetmacro{\ewaY}{2*sqrt(\ewaDet*\ewaA)/\ewaDet}
	\pgfmathsetmacro{\myw}{(\ewaX+1)*#1}
	\pgfmathsetmacro{\myh}{(\ewaY+1)*#1}
	\pgfmathparse{\nx < 0}
	\ifthenelse{\equal{\pgfmathresult}{1}}{
		\pgfmathsetmacro{\mycx}{#1*0.5*(\ewaX-1)}
	}{
		\pgfmathsetmacro{\mycx}{#1*0.5*(1-\ewaX)}
	}%
	\pgfmathparse{\ny < 0}
	\ifthenelse{\equal{\pgfmathresult}{1}}{
		\pgfmathsetmacro{\mycy}{#1*0.5*(\ewaY-1)}
	}{
		\pgfmathsetmacro{\mycy}{#1*0.5*(1-\ewaY)}
	}%
	\node[tdplot_screen_coords] at (\mycx, \mycy, 0) {\includegraphics[width=\myw cm,height=\myh cm]{#4}};
\end{scope}
}

\tikzset{
	mytdplotSphere/.style={thin,black!30,tdplotCsBack/.style={draw=none}},
	mytdplotSphereUpper/.style={mytdplotSphere},
	mytdplotSphereBase/.style={mytdplotSphere},
	mytdplotSphereTheta/.style={mytdplotSphere,opacity=0.5},
	mytdplotSpherePhi/.style={mytdplotSphere,opacity=0.5},
	mytdplotSphereEquator/.style={mytdplotSphere},
}
\newcommand{\mytdplotdrawsphere}[2][]{%
\begin{scope}[#1]
	\foreach \i in {1,...,4}{
		\tdplotCsDrawLonCircle[mytdplotSphereTheta]{#2}{(180*\i+90)/4}
	}
	\foreach \i in {-3,-2,-1,1,2,3}{
		\tdplotCsDrawLatCircle[mytdplotSpherePhi]{#2}{90*\i/4}
	}
	\tdplotCsDrawLatCircle[mytdplotSphereEquator]{#2}{0}
	\begin{scope}[tdplot_screen_coords]
		\pgfmathsetmacro{\rx}{#2}
		\pgfmathsetmacro{\ry}{\rx*cos(\tdplotmaintheta)}
		\draw[mytdplotSphere] (0:\rx) arc (0:360:\rx);
	\end{scope}
\end{scope}
}

\tikzset{
	mytdplotAxis/.style={thick,-Latex,line cap=round},
	mytdplotAxisMinor/.style={thick,line cap=round,line join=round},
}

\newcommand{\mytdplotdrawdeliotgrid}[2][]{%
\begin{scope}[#1]
	\pgfmathsetmacro{\myr}{#2}
	\pgfmathsetmacro{\myrext}{#2+0.5}
	\draw[mytdplotAxisMinor] (0,0,0) -- ++(\myr,0);
	\draw[mytdplotAxisMinor] (0,0,0) -- ++(0,\myr);
	\foreach \myi in {1,...,\myr}{
		\draw[mytdplotAxisMinor] (\myi,0) arc (0:90:\myi);
		\ifthenelse{\equal{\myr}{\myi}}{}{
			\pgfmathsetmacro{\myn}{2^(\myi-1)}
			\foreach \myj in {1,...,\myn}{
				\pgfmathsetmacro{\myphi}{(\myj-0.5)/\myn*90}
				\draw[mytdplotAxisMinor] (\myphi:\myi) -- (\myphi:\myr);
			}
		}
	}
\end{scope}
}

\def\mytdplotdefaultcoordsONV{
	\coordinate (O) at (0,0,0);
	\mytdplotsetcoordclippedext{N}{\r}{\thetaN}{\phiN}
	\mytdplotsetcoordclippedext{V}{\r}{\thetaV}{\phiV}
	\mytdplotsetcoordclippedext{R}{\r}{\thetaR}{\phiR}
	
	\begin{scope}[tdplot_screen_coords]
		\path (O) ++ (-\r, 0, 0) coordinate (Owest);
		\path (O) ++ (\r, 0, 0) coordinate (Oeast);
		\pgfmathsetmacro{\myy}{-\r*cos(\tdplotmaintheta)}
		\path (O) ++ (0, \myy, 0) coordinate (Osouth);
	\end{scope}
}
\tikzset{
	dirStyle/.style={
		thick,-Latex,font=\small,
		labelStyle/.style={above=0.2em},
	},
	dirVstyle/.style={
		dirStyle,
		labelStyle/.style={above left},
	},
	dirRstyle/.style={
		dirStyle,
		labelStyle/.style={above right},
	},
	dirNstyle/.style={
		dirStyle,
		labelStyle/.style={above=0.2em},
	},
}

\def\mytdplotdefaultdirsR{
	\draw[dirRstyle] (Rboundary) -- (Rext) node[labelStyle] {$\dirR$};
}

\def\mytdplotdefaultdirsNV{
	\draw[dirVstyle] (Vboundary) -- (Vext) node[labelStyle] {$\dirV$};
	\draw[dirNstyle] (Nboundary) -- (Next) node[labelStyle] {$\dirN$};
}

\tikzset{
	unitStyle/.style={
		at=(Osouth-|Oeast),
		font=\footnotesize
	}
}

\tdplotsetmaincoords{70}{45} 
\pgfmathsetmacro{\r}{2}
\pgfmathsetmacro{\thetaN}{0}
\pgfmathsetmacro{\phiN}{90}
\pgfmathsetmacro{\thetaV}{50}
\pgfmathsetmacro{\phiV}{180}
\pgfmathsetmacro{\thetaR}{\thetaV}
\pgfmathsetmacro{\phiR}{\phiV-180}
\pgfmathsetmacro{\thetaL}{60}
\pgfmathsetmacro{\phiL}{15}

\begin{document}

\newcommand{\imgpathteaser}[2]{assets/teaser/#1/#1_logn#2.png}
\def\teaserheight{6.31cm}
\def\teasersepscale{0.012}
\pgfmathparse{\teaserheight*\teasersepscale}
\edef\teasersep{\pgfmathresult pt}
\def\teaserdresslogn{7.0}
\def\teasercarlogn{16.0}
\def\teaserornamentslogn{14.0}
%
\newcommand{\maketrim}[6]{
	\def\triml{#1}
	\def\trimb{\numexpr#6-#4-#2\relax}
	\def\trimr{\numexpr#5-#3-#1\relax}
	\def\trimt{#2}
}
\teaser{
	\centering
	\begin{tikzpicture}[node distance=\teasersep, every node/.style={inner sep=0, outer sep=0}]
		\tikzset{
			imgnode/.style={
				draw, line width=1.5pt,line join=round
			},
			insetnode/.style={
				draw, line width=3pt,line join=round
			},
			insetOverlay/.style={
				line width=1pt, line join=round,
			},
			colorA/.style={
				set1violet
			},
			colorB/.style={
				set1blue
			},
			colorC/.style={
				set1green
			},
			colorD/.style={
				set1orange
			},
		}
		\node[imgnode] (carmain) {\includegraphics[height=\teaserheight,trim=400 130 100 30,clip]{\imgpathteaser{car}{\teasercarlogn}}};
		\node[imgnode,left=of carmain] (dressmain) {\includegraphics[height=\teaserheight]{\imgpathteaser{dress}{\teaserdresslogn}}};
		\node[imgnode,right=of carmain.south east, anchor=south west] (ornamentsmain) {\includegraphics[height=\dimexpr \teaserheight*3/4-\teasersep/4 \relax,trim=0 5 0 0,clip]{\imgpathteaser{ornaments}{\teaserornamentslogn}}};

		\begin{scope}[
			x={($(dressmain.north east)-(dressmain.north west)$)},
			y={($(dressmain.south west)-(dressmain.north west)$)},
			shift={(dressmain.north west)},
			xscale=0.142857143, 
			yscale=0.0625       
			]
			\draw[insetOverlay,colorA] (1.58,3.47) rectangle ++(2,2);
		\end{scope}

		\begin{scope}[
			x={($(carmain.north east)-(carmain.north west)$)},
			y={($(carmain.south west)-(carmain.north west)$)},
			shift={(carmain.north west)},
			xscale=0.052631579, 
			yscale=0.06944444444444445
			]
			\draw[insetOverlay,colorB] (8.7,6.31) rectangle ++(2,2);
		\end{scope}
		
		\begin{scope}[
			x={($(ornamentsmain.north east)-(ornamentsmain.north west)$)},
			y={($(ornamentsmain.south west)-(ornamentsmain.north west)$)},
			shift={(ornamentsmain.north west)},
			xscale=0.0625,     
			yscale=0.083682008 
			]
			\draw[insetOverlay,colorC] (2.82,4.12) rectangle ++(2,2);
			\draw[insetOverlay,colorD] (9.64,6.05) rectangle ++(2,2);
		\end{scope}

		\maketrim{158}{347}{200}{200}{700}{1600}
		\node[right=of carmain.north east,anchor=north west, minimum size=\dimexpr\teaserheight/4-\teasersep*3/4\relax] (dresscrop) {};
		\node[insetnode,colorA,at=(dresscrop)]
		{\includegraphics[height=\dimexpr\teaserheight/4-\teasersep*3/2\relax,trim={{\triml} {\trimb} {\trimr} {\trimt}},clip]{\imgpathteaser{dress}{\teaserdresslogn}}};

		\maketrim{1270}{661}{200}{200}{2400}{1600}
		\node[right=of dresscrop, minimum size=\dimexpr\teaserheight/4-\teasersep*3/4\relax] (carcrop) {};
		\node[insetnode,colorB,at=(carcrop)]
		{\includegraphics[height=\dimexpr\teaserheight/4-\teasersep*3/2\relax,trim={{\triml} {\trimb} {\trimr} {\trimt}},clip]{\imgpathteaser{car}{\teasercarlogn}}};

		\maketrim{282}{412}{200}{200}{1600}{1200}
		\node[right=of carcrop, minimum size=\dimexpr\teaserheight/4-\teasersep*3/4\relax] (ornamentcrop1) {};
		\node[insetnode,colorC,at=(ornamentcrop1)]
		{\includegraphics[height=\dimexpr\teaserheight/4-\teasersep*3/2\relax,trim={{\triml} {\trimb} {\trimr} {\trimt}},clip]{\imgpathteaser{ornaments}{\teaserornamentslogn}}};

		\maketrim{964}{605}{200}{200}{1600}{1200}
		\node[right=of ornamentcrop1, minimum size=\dimexpr\teaserheight/4-\teasersep*3/4\relax] (ornamentcrop2) {};
		\node[insetnode,colorD,at=(ornamentcrop2)]
		{\includegraphics[height=\dimexpr\teaserheight/4-\teasersep*3/2\relax,trim={{\triml} {\trimb} {\trimr} {\trimt}},clip]{\imgpathteaser{ornaments}{\teaserornamentslogn}}};

	\end{tikzpicture}

	%
	%
	%
	\caption{
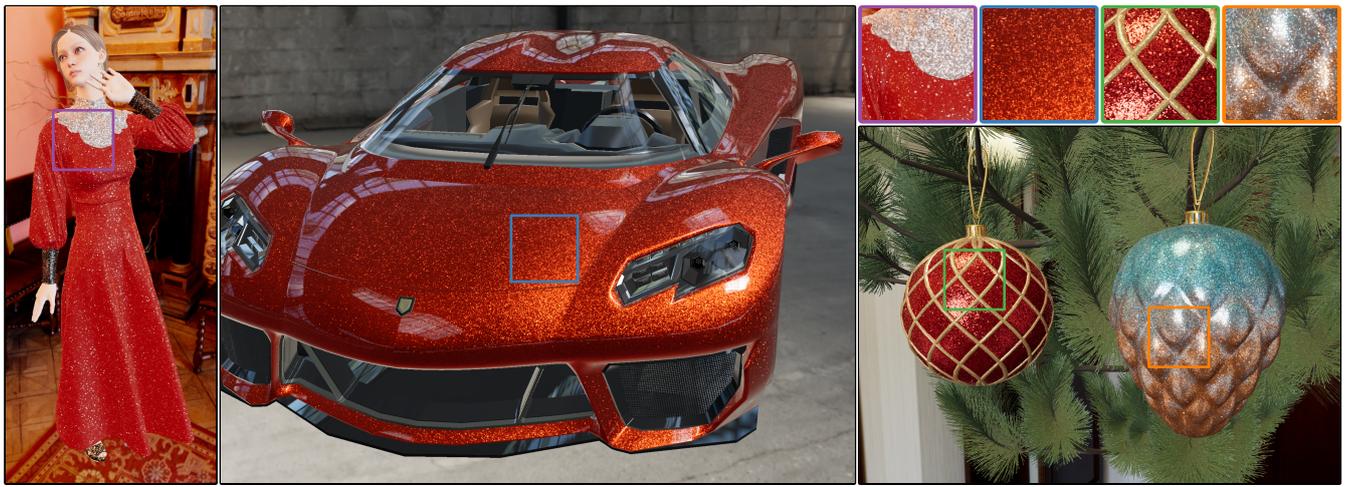
		Various glittery materials are rendered with our real-time image-based lighting approximation.
		From left to right:
		We show a woman wearing a glittering dress, a car with metallic paint, and glittery ornamental balls.
		Insets are shown in the top right.
	}
	\label{fig:teaser}
}


\maketitle

\begin{abstract}
	Image-based lighting is a widely used technique to reproduce shading under real-world lighting conditions, especially in real-time rendering applications.
	A particularly challenging scenario involves materials exhibiting a sparkling or glittering appearance, caused by discrete microfacets scattered across their surface.
	In this paper, we propose an efficient approximation for image-based lighting of glints, enabling fully dynamic material properties and environment maps.
	Our novel approach is grounded in real-time glint rendering under area light illumination and employs standard environment map filtering techniques.
	Crucially, our environment map filtering process is sufficiently fast to be executed on a per-frame basis.
	Our method assumes that the environment map is partitioned into few homogeneous regions of constant radiance.
	By filtering the corresponding indicator functions with the normal distribution function, we obtain the probabilities for individual microfacets to reflect light from each region.
	During shading, these probabilities are utilized to hierarchically sample a multinomial distribution, facilitated by our novel dual-gated Gaussian approximation of binomial distributions.
	We validate that our real-time approximation is close to ground-truth renderings for a range of material properties and lighting conditions, and demonstrate robust and stable performance, with little overhead over rendering glints from a single directional light.
	Compared to rendering smooth materials without glints, our approach requires twice as much memory to store the prefiltered environment map.
	%

\ccsdesc[500]{Computing methodologies~Rendering}
\ccsdesc[500]{Computing methodologies~Reflectance modeling}

\printccsdesc
\end{abstract}

%
\section{Introduction}
\label{sec:introduction}
Since time immemorial, humankind has been attracted towards shiny and glittery objects \cite{leddy1997sparkle}.
In this work, we are particularly interested in the sparkling effect of glints, which are usually the result of mesoscopic mirror-like surface features.
Examples range from metallic car paints in the automotive industry to makeup and fancy dresses in the fashion industry, but also arises as a natural phenomenon in snow, stones, or when light is reflected from the waves of an ocean.
Reproducing these effects accurately in virtual worlds is critical to the immersion and authenticity.
%
%
Consistently generating new images within a tight time budget is of equal importance, especially in interactive applications with low latency requirements like video games.
%

%
%
%
In order to provide authentic real-world illumination to synthetic scenes, image-based lighting is commonly employed across the board.
Instead of manually placing light sources, an environment map is captured, which records the amount of light incident from each direction.
%
%
%
It is frequently used in interactive applications to model distant illumination from the environment, and to account for (one-bounce) indirect illumination via reflection captures.


%
Reproducing the appearance of virtual shiny objects has been rigorously studied since the early days of computer graphics \cite{blinn1977models}, culminating in microfacet theory \cite{cook1982reflectance}.
The sparkling effect of discrete microfacets has been investigated and formalized much later \cite{jakob2014discrete,yan2014rendering} and is much more expensive to simulate.
Glints are challenging to render, due to their high-frequency behavior, both spatially and with respect to viewing and lighting conditions.
The problem of real-time glint rendering with point and area light sources has been addressed with practical solutions \cite{chermain2020procedural,deliot2023real,kneiphof2024real}.
%
%
Our main contribution is an extension of previous real-time glint rendering techniques \cite{deliot2023real,kneiphof2024real} to support image-based lighting.
\cref{fig:teaser} showcases a selection of renderings generated with our method.
Nevertheless, we also address some artifacts that occur under point or area light illumination.
Previous work \cite{jakob2014discrete,kneiphof2024real} has demonstrated that the size of light sources has a direct impact on the appearance of glittery surfaces.
Small and large light sources emitting identical radiant power might lead to a similar reflected radiance on smooth surfaces.
On a glittery surface however, incident light distributed over a small solid angle leads to sparse but bright glints, since only a small fraction of the microfacets will be correctly oriented for reflection.
For light sources covering a large solid angle, microfacets will have a higher probability for reflection, but each one transports a lower amount of radiance.
Our method links the cheap and readily available real-time image-based lighting solutions \cite{karis2013real,lagarde2014moving} with an existing counting model for glint rendering \cite{deliot2023real}.
Analogous to real-time glint rendering under area light illumination \cite{kneiphof2024real}, we consider the solid angle area of regions in the environment map that emit similar radiance.
We assume that the environment map is partitioned into few homogeneous but arbitrarily shaped regions of constant radiance.
Standard environment map filtering is employed to compute the probabilities for a microfacet to be oriented for reflection from each region.
%
The microfacets inside of a given pixel footprint are assigned to the regions by sampling the approximation of a multinomial distribution during shading.
To facilitate the sampling of the multinomial distribution in constant time, we propose a new dual-gated Gaussian approximation for binomial distribution sampling.
Hierarchically sampling the binomial approximations leads to the final multinomial sampling.
Each microfacet then reflects the amount of radiance that is associated with the respective region of incoming light.
The time required to shade a pixel with our technique is constant, independent of the microfacet density, surface roughness or environment map, and only requires a few channels of additional environment map data to be prefiltered in real-time applications.
There has been an earlier attempt at real-time image-based lighting of glints by Wang et al. \cite{wang2020real}.
The increased runtime due to a quad tree traversal and explicit summation over glints is easily addressed by the anisotropic grid and gated Gaussian approximation for sampling binomial distributions \cite{deliot2023real}, that also form the basis of our work.
The main difference to our method is a difference in choice of the probability for a microfacet to reflect light from the environment map, which they compute independently of the contents of the environment map.
As we show in our results, this leads to inferior visual quality.
Our main contributions are:
\begin{itemize}
	\item Formulation of glint rendering for multiple non-overlapping area lights, given their individual domains of incident light and combined reflected radiance from an equivalent smooth surface, and the application to environment maps.
	\item A dual-gated Gaussian approximation for improved symmetry during sampling of binomial distributions.
	\item Generalization of sampling binomial distributions to real-valued trial counts, especially small values.
\end{itemize}
The core assumptions of our real-time approximation are:
\begin{itemize}
	\item Both the lobe of the bi-directional reflection distribution function (BRDF) and the microfacet normal distribution function (NDF) are well approximated with the lobe that is commonly used to prefilter environment maps for real-time image-based lighting.
	Only their amplitude differs.
	\item The glints only modulate the brightness of the outgoing radiance (i.e. we assume radiance to be scalar-valued in our glint computation), and the chromatic information is taken from an equivalent smooth BRDF.
	\item The environment map is well approximated, if partitioned into few homogeneous regions of constant radiance.
	%
\end{itemize}
%


%
\section{Related Work}
\label{sec:related_work}
In the following, we review previous work on glint rendering, as well as real-time image-based lighting.
Zhu et al. \cite{zhu2022recent} provide an overview over recent offline methods for glint rendering.
Most approaches can be classified as being stochastic \cite{jakob2014discrete}, or using high-resolution normal maps \cite{yan2014rendering}.
\paragraph{Stochastic Glints}
Our real-time approach relates to the stochastic glint category, which was pioneered by \cite{jakob2014discrete} for offline rendering, where discrete microfacets are randomly distributed in the spatio-angular domain.
This approach serves as the basis for several improvements \cite{atanasov2016practical, wang2018fast}.
The first real-time capable approach in this line of work handles point and environment illumination using a prefiltering approach\cite{wang2020real}.
One limitations is that the probability for a glint to appear under image-based lighting is independent of the contents of the environment map.
This is directly addressed by our method.

\paragraph{Counting Models}
All the stochastic methods discussed in the previous paragraph procedurally instantiate microfacets in the spatio-angular domain, which has to be searched for evaluation.
In contrast, we replace the spatio-angular search for discrete microfacets with a counting model based on an approximation of binomial distributions which is naturally fast enough for real-time applications \cite{zirr2016real,deliot2023real,kneiphof2024real}.
The approach of Deliot and Belcour \cite{deliot2023real} has been extended for area lights \cite{kneiphof2024real}, which forms the basis of our real-time image-based lighting approach.
In general, importance sampling of glints \cite{chermain2021importance} is a straight forward approach to solve image-based lighting.
For low microfacet densities, explicitly sampling a few reflections per pixel is reasonable, but it does not scale to surface with densely packed microfacets on the surface.
Our method has constant cost and scales seamlessly from very low to very high microfacet densities.
\paragraph{High Resolution Normal Maps}
Alternatively, a high resolution normal map can be employed to yield a glittery appearance \cite{yan2014rendering, yan2016position}.
This approach is sped up by a more efficient integration of the NDF over pixel footprint and light source \cite{gamboa2018scalable, atanasov2021multiscale, deng2022constant}, and the storage cost is reduced using procedural normal maps \cite{wang2020example} or an example-based approach \cite{xing2024tiny}.
Using steeply sloped normal maps requires to consider multiple scattering \cite{chermain2019glint}.
Neural and differentiable rendering techniques have also been employed to synthesize glittery appearance in this context\cite{kuznetsov2019learning,fan2022efficient,shah2024neural}.
If the geometric features in the normal map become small enough, chromatic variations manifest due to diffraction\cite{yan2018rendering}.
The subjective optical speckle due to diffraction on random rough surfaces was investigated in more detail by Steinberg and Yan\cite{steinberg2022rendering}.
In our real-time image-based lighting approach, we do not consider chromatic variations of glints.
%

%
%
%

%
\paragraph{Real-Time Glints}
Early procedural methods for real-time rendering produce highlights by intersecting the surface with a world-space grid\cite{bowles2015sparkly, wang2016robust}.
Multiple real-time approaches fall into the discrete stochastic line of work \cite{wang2020real,chermain2020procedural,zirr2016real,deliot2023real,kneiphof2024real}, while other fast approaches employ normal map pre-filtering \cite{chermain2021real,tan2022real,xing2024tiny}.
\paragraph{Car Paint Models}
Metallic car paints are well known for their sparkling appearance, caused by metallic flakes embedded in the substrate\cite{ershov1999simulation,ershov2001rendering}.
Domain-specific data-driven representations have been developed for flakes\cite{rump2008photo, rump2009efficient,golla2017efficient}.
Another effect often observed in metallic paints are iridescent color shifts\cite{belcour2017practical,kneiphof2019real,kneiphof2022real}.
Both effects are unified in a single offline rendering framework by Guo et al. \cite{guo2018physically}.
Properly supporting chromatic variations of individual glints in real-time is still a challenging endeavor.
%

%
%
%

%
%
%

%
\paragraph{Real-time Image-based Lighting}
For low-frequent reflection models, few spherical harmonics coefficients are sufficient to encode the filtered environment map \cite{ramamoorthi2001efficient}.
Our glint model assumes microfacet distributions resulting in glossy surfaces.
Current real-time image-based lighting techniques \cite{karis2013real,lagarde2014moving,mcauley2015rendering} are based on prefiltering the environment map \cite{kautz2000unified}.
Some improvements have been proposed, for example faster environment map filtering \cite{manson2016fast},
or correction of energy loss due to multiple-scattering \cite{fdez2019multiple}.
We do not consider multiple scattering when dealing with glints, as they are overwhelmingly caused by single scattering.
Cocco et al. \cite{cocco2024anisotropic} introduce an anisotropic extension to real-time image-based lighting, which is orthogonal to our work and can be directly used for anisotropic microfacet distributions.
Occlusion is typically neglected in real-time image-based lighting.
However, it can be taken into account using a stochastic occlusion estimate \cite{heitz2018combining}.
%
For glints however, it is not sufficient to simply attenuate the resulting appearance, since the occlusion affects the probability of individual glints being present.
%
For shadow mapping, approximations of environment maps using point lights \cite{annen2007convolution} or area lights \cite{kneiphof2017real} have been tried.
However, a large number of light sources is required and the approximation leads to artifacts on low-roughness surfaces.
In our work, we do make use of an area light interpretation of the environment map, but we do not constrain the shape of the area lights and require them to partition the whole hemisphere.
This allows us to only consider a low number of light sources in our implementation while simultaneously avoiding the artifacts on low-roughness surfaces.
%



%
\section{Background}
\label{sec:preliminaries}
Our technique is built around the theoretical framework of discrete normal distributions \cite{jakob2014discrete,kneiphof2024real}, as well as the commonly used real-time image-based lighting \cite{karis2013real,lagarde2014moving} for smooth reflection models.
%


%
\subsection{Rendering Equation and the Pixel Footprint}
\label{sec:rendering}
In accordance with microfacet theory \cite{cook1982reflectance}, the bi-directional reflectance distribution function (BRDF) describes the light transport from direction $\dirI$ to $\dirO$ as
\begin{align}
	\brdf = \frac{\fresnel \cdot \geom \cdot \ndf}{4 \cdot \ndotv \cdot \ndotl},
\end{align}
where $\sprod{\cdot}{\cdot}$ denotes the dot product, $\dirN$ is the macroscopic surface normal, $\dirH := \nicefrac{(\dirI+\dirO)}{\|\dirI+\dirO\|}$ is the halfway vector, $\fresnel$ is the Fresnel reflectance term for perfectly smooth microfacets, $\geom$ is the geometric shadowing and masking term, and $\ndf$ is the normal distribution function (NDF) parameterized by a roughness value $\alpha$ which is omitted for brevity.
Since microfacets are perfectly smooth in our model, the halfway vector $\dirH$ is identical to the microfacet normal $\dirM$ that reflects light from $\dirI$ to $\dirO$.
For brevity, we define $\brdfnol := \brdf \cdot \ndotl$.
The rendering equation \cite{kajiya1986rendering} integrates the BRDF over incoming directions to get the outgoing radiance of a smooth surface
\begin{align}
	\label{eq:rendering}
	\Lo = \int_\domainHemi \brdfnol \cdot \Li \dx{\dirI},
\end{align}
where $\Li$ is the incoming radiance, e.g. from the environment map.
This holds for an infinitesimally small shading point $\pos$ on the surface.
For our glint model however, we need to take the surface footprint $\pxFoot$ of each pixel into account.
\begin{align}
	\LoFoot = \int_\domainHemi \frac{\fresnel \cdot \geom \cdot \ndfFoot}{4\cdot\ndotv} \cdot \Li \dx{\dirI},
\end{align}
where all terms of the BRDF are assumed constant within $\pxFoot$.
Only the NDF $\ndfFoot$ is chosen to depend on the shape of the pixel footprint, which we define in the next section as the discrete NDF responsible for glints.
%


%
\subsection{Discrete Normal Distribution}
\label{sec:discretendf}
Now let us introduce the discrete normal distribution model that is responsible for the glint appearance.
The formulation is borrowed from the theory previously used for area lights \cite{kneiphof2024real}.
The smooth BRDF model is the result of a continuous distribution of infinitely many and infinitesimally small microfacets.
If the reflectance is modeled using a finite number of microfacets instead, glints start to appear.
When the number of microfacets is increased in the glint model, its appearance is expected to converge to that of the smooth model.
The continuous NDF $\ndf$ measures the density of differential microfacet area per differential surface area, having any given orientation $\dirH$.
The discrete microfacets are assumed to be identically sized, irrespective of their orientation.
Therefore, we want them to be distributed proportionally to the unprojected microfacet area measured by $\ndf$.
Since the surface is assumed not to contain any holes, only the microfacet area projected onto the surface integrates to exactly $1$.
In order to turn $\ndf$ into a probability density function, it needs to normalized by the total microfacet area per unit surface area $\intDHemi := \intDHemiLong \geq 1$.
%


%
%
Let there be $\Ntotal$ microfacets i.i.d. per unit surface patch
, each producing a Dirac-delta reflection.
For $1 \leq k \leq \Ntotal$, their positions $\posk \propto \mathcal{U}[0,1]^2$ are uniformly distributed, and their orientations $\dirHk \propto \nicefrac{\ndfArgs{\dirHk}}{\intDHemi}$ are distributed proportionally to the underlying smooth model.
For a given pixel footprint $\pxFoot$, the discrete NDF model is then defined as
\begin{align}
	\ndfFoot &:= \intDHemi \cdot \frac{1}{\Ntotal\cdot|\pxFoot|}
	\int_\pxFoot \sum_{k=1}^\Ntotal \delta_\pos(\pos - \posk) \cdot \delta_\dirH(\dirH - \dirHk) \dx{\pos} \\
	&= \intDHemi \cdot \frac{1}{\Ex{\Nfoot}} \cdot \sum_{k=1}^{\Nfoot} \delta_\dirH\left(\dirH - \dirHk\right),
	\label{eq:discrete_ndf}
\end{align}
where $\delta_\pos$ and $\delta_\dirH$ are Dirac-delta functions with respect to surface area and solid angle measure.
In \cref{eq:discrete_ndf} only the microfacets inside of the pixel footprint are considered.
Assuming i.i.d. microfacets, the actual number of microfacets inside of the pixel footprint $\Nfoot = b(N, |\pxFoot|)$ is the result of a counting process, assuming $|\pxFoot| \leq 1$.
The expected number of microfacets inside of the pixel footprint is $\Ex{\Nfoot} = N\cdot|\pxFoot|$, the expected value of the binomial distribution.
In practice, the microfacets are assumed to be evenly stratified across the surface \cite{deliot2023real}, resulting in $\Nfoot = N \cdot |\pxFoot| = \Ex{\Nfoot}$, sparing the sampling of a binomial distribution or the evaluation of a more expensive counting mechanism \cite{wang2020real}.
In the following, we assume an arbitrary area light source with constant radiance $\totalLi$, covering the domain of incoming directions $\domainI$.
By element-wise bisection of $\domainI$ and $\dirO$, the domain of microfacet orientations $\domainH = \{\dirH = \nicefrac{(\dirO + \dirI)}{\|\dirI + \dirO\|} \mid \dirI \in \domainI \}$ responsible for a reflection from the light source towards the observer is obtained.
When computing the reflected radiance from a glittery surface, we want to count the number of microfacets that reflect light form the area light towards the observer.
This is expressed as an integration of the discrete NDF over the domain of reflecting microfacet orientations $\domainH$, defined by the incoming light directions $\domainI$.
%
%
\begin{align}
	\int_\domainH \ndfFoot \dx{\dirH} = \frac{\intDHemi}{\Ex{\Nfoot}} \cdot
	\sum_{k=1}^{\Nfoot} \mathbf{1}_\domainH\left(\dirHk\right) \dx{\dirH},
\end{align}
where $\mathbf{1}_\domainH(\dirHk)$ is the indicator function that yields $1$ if $\dirHk \in \domainH$ and $0$ otherwise.
The sum can now be represented using a binomial distribution as
\begin{align}
	\label{eq:binomial_intd}
	\int_\domainH \ndfFoot \dx{\dirH} = \intDHemi \cdot \frac{b(\Nfoot, \probH)}{\Ex{\Nfoot}},
\end{align}
where $\probH$ is the probability for any microfacet to be correctly oriented for reflection, is defined as the fraction of correctly oriented microfacet surface area and the total microfacet area according to the underlying smooth NDF
\begin{align}
	\probH := \frac{\intDLong}{\intDHemiLong} = \frac{\intD}{\intDHemi}
	\label{eq:probh_def}
\end{align}
In practice we do not consider $\domainH$ explicitly, and only work with $\domainI$.
Therefore $\intD$ is rewritten using a change of variables with respect to $\domainI$ as
\begin{align}
	\intD = \int_\domainI \frac{\ndf}{4\cdot\hdotl} \dx{\dirI}.
\end{align}
For shorthand notation, we define $\ndfhol := \nicefrac{\ndf}{(4\cdot\hdotl)}$ analogous to $\brdfnol$.
%


%
\subsection{Binomial Sampling with Anisotropic Grids}
\label{sec:anisogrid}
To attain temporal stability, the number of microfacets inside of the pixel footprint are distributed on a multi-dimensional grid \cite{deliot2023real}.
The shape of the pixel footprint is parameterized by its length, anisotropy factor (major axis divided by minor axis) and orientation on the surface.
The number of microfacets $\Nfoot$ is distributed to the adjacent vertices inside of a tetrahedral cell in that grid.
Individual binomial distributions (or in our case multinomial distributions, \cref{sec:theory}) are sampled at the vertices of the grid, and the resulting sample counts are summed again.
The surface is also subdivided into a triangle grid \cite{deliot2019procedural} in the space of uv coordinates, where the outcomes of the binomial samplings at the grid vertices are linearly interpolated using the shading position $\pos$.
For directional (and area) light sources, the same strategy is used for an interpolation using the halfway vector $\dirH$.
With image-based lighting however, we do not have access to a meaningful halfway vector.
Therefore, we omit the distribution over the halfway vector, leading to a lower-dimensional grid that disregards $\dirH$.
Sampling the binomial distribution in \cref{eq:binomial_intd} for a given pixel footprint is thus realized as a weighted sum
\begin{equation}
	b(\Nfoot, \probH) = \sum_{i \in \mathcal{I}} w_i \cdot b_{\theta_i}(N_i, \probH),
	\label{eq:anisogrid}
\end{equation}
%
where $w_i$ are the interpolation weights of the grid vertices, $\theta_i$ are random seed stored at the grid vertices, and $N_i$ are the number of microfacets at the grid vertices.
Since we will use the distribution of microfacets on the anisotropic grid as a black box, please refer to the original work \cite{deliot2023real} for more details.
%


\iffigsinline

\begin{figure}
	\centering
	\pgfmathsetmacro{\r}{1.3}
	\newcommand{\hemiimgpath}[1]{assets/hemisphereplot/output/#1.png}
	\tikzset{
		descNode/.style={
			font=\small
		},
		hemisphereOutline/.style={
			draw,
			line width=0.75pt,
			black
		}
	}
	\subfloat[BRDF]{%
	\label{fig:hemisphereplot:brdf}%
	\begin{tikzpicture}[every node/.style={inner sep=0, outer sep=0}]
		\begin{scope}[tdplot_main_coords]
			\mytdplotdefaultcoordsONV
			\mytdplotdrawhemispherebg{\r}{\hemiimgpath{brdf_light}}
			\mytdplotdrawsphere{\r}
			\mytdplotdrawhemisphereoutline[hemisphereOutline]{\r}
			\mytdplotdefaultdirsNV
		\end{scope}
	\end{tikzpicture}%
	}%
	\subfloat[NDF]{%
	\label{fig:hemisphereplot:ndf}%
	\begin{tikzpicture}[every node/.style={inner sep=0, outer sep=0}]
		\begin{scope}[tdplot_main_coords]
			\mytdplotdefaultcoordsONV
			\mytdplotdrawhemispherebg{\r}{\hemiimgpath{ndf_light}}
			\mytdplotdrawsphere{\r}
			\mytdplotdrawhemisphereoutline[hemisphereOutline]{\r}
			\mytdplotdefaultdirsNV
		\end{scope}
	\end{tikzpicture}%
	}%
	\subfloat[IBL Kernel]{%
	\label{fig:hemisphereplot:ibl}%
	\begin{tikzpicture}[every node/.style={inner sep=0, outer sep=0}]
		\begin{scope}[tdplot_main_coords]
			\mytdplotdefaultcoordsONV
			\mytdplotdrawhemispherebgrot{\r}{\thetaV}{\phiV-180}{\hemiimgpath{ibl_brdf_light}}
			\mytdplotdrawsphere{\r}
			\mytdplotdrawhemisphereoutlinerot[hemisphereOutline]{\r}{\thetaV}{\phiV-180}
			\mytdplotdefaultdirsR
		\end{scope}
	\end{tikzpicture}
	}%
	\caption{
		Illustration of lobe shapes on a rough surface with $\alpha=0.5$ and viewing angle $\theta_o = \ang{50}$.
		(\subref{fig:hemisphereplot:brdf}) shows the smooth BRDF $\brdfnol$, (\subref{fig:hemisphereplot:ndf}) shows the underlying NDF $\ndfhol$ and (\subref{fig:hemisphereplot:ibl}) shows the corresponding kernel $\iblKernel$ used to filter the environment map, which assumes $\dirN = \dirV = \dirR$.
	}
	\label{fig:hemisphereplot}
\end{figure}
\fi

\subsection{Real-time Image-based Lighting}
\label{sec:rtibl}
We briefly recapitulate how image-based lighting is commonly implemented in real-time applications \cite{karis2013real, lagarde2014moving}.
The environment map is prefiltered with kernels for different roughness values.
During shading the prefiltered environment map is used to approximate the the outgoing radiance after reflection.
Besides the radiance, our method augments the environment map with additional data, which is prefiltered using the same technique.
The integral in the rendering equation (\cref{eq:rendering}) is approximated using a split-sum approximation \cite{karis2013real}.
The BRDF is integrated disregarding the illumination, yielding the directional albedo $\brdfFGD$ in the first integral,
and the environment map is filtered for given reflection direction $\dirR$ in the second integral.
\begin{align}
	\Lo &= \underbrace{\int_\domainI \brdfnol \dx{\dirI}}_{=:\brdfFGD} \cdot \int_\domainI \iblKernel \cdot \Li \dx{\dirI},
	\label{eq:rtibl}
\end{align}
where $\iblKernel \propto \brdfnolArgs{\dirI,\dirR}$ is a probability distribution on the light directions $\dirI$ that is used to prefilter the environment map, under the assumption that $\dirN = \dirV = \dirR$, illustrated in \cref{fig:hemisphereplot}.

\hide{
	Compare directional albedo $\brdfFGD$ with
	\enquote{total NDF visible blabla} $\ndfFGD$,
	directional integrated NDF, disregarding shadowing and masking.

	Reference the directional albedo figure when added
}



\iffigsinline\input{source/figure/radiancelevels.tex}\fi

%
\section{Image-based Lighting as Multiple Area Lights}
\label{sec:theory}

This section introduces the formulation of glint rendering under image-based lighting as illumination from multiple area lights, based on the discrete NDF model (\cref{sec:discretendf}).
For simplicity, we assume that the incoming radiance from the environment map is scalar and varies only in brightness.
Chromatic variations due to the incident light are modeled only by the equivalent smooth BRDF.
The environment map is assumed to be partitioned into a few area lights of constant brightness (c.f. \cref{fig:radiancelevels} (top)).
During shading, the probability for a microfacet to reflect light from each of the area lights is estimated.
%
The outgoing radiance for the glint model ends up as the outgoing radiance of the smooth model, modulated by a weighted multinomial sample divided by its identically weighted expected value.
The practical aspects of this formulation are detailed in \cref{sec:practice}.
%


%
\paragraph{Light Partition}
The sphere of incoming light directions is assumed to be partitioned into $K$ arbitrary area lights.
Each light emits constant radiance $\Lik$ and covers the incoming light directions $\domainIk \subseteq \domainSph$.
Furthermore, we define functions $\wkSymb : \domainSph \rightarrow \{0, 1\}$ with $\sum_{k=1}^{K}\wk = 1$, that indicate which light source the direction $\dirI$ belongs to.
The incoming radiance from the environment map is then written as the combined incoming radiance from all the area lights
\begin{align}
	\label{eq:liksum}
	\Li = \sum_{k=1}^K \wk \cdot \Lik.
\end{align}
We found that $K\in\{4,8\}$ is sufficient for real-world environments.
Although our method generalizes to larger $K$, this would only benefit extreme dynamic range illumination, unlikely to be encountered in practice.
The partitioning of the incoming light directions is relaxed into a convex combination in \cref{sec:practice}, turning $\wkSymb : \domainSph \rightarrow [0, 1]$ into fuzzy indicator functions.
Theoretically, this definition holds for color-valued radiance.
In practice however, we only consider the brightness when determining $\Lik$ and the associated $\wk$ in \cref{sec:practice} for simplicity.
\paragraph{Mean Reflectance}
A central assumption for rendering glints illuminated by area lights in real time is that the reflectance and geometric masking shadowing terms are assumed constant for all participating incoming light directions $\dirI \in \domainI$ \cite{kneiphof2024real}, i.e. $\brdfnol \sim \ndfhol$ is assumed locally given $\theta_o$.
This is carried over for image-based lighting in a slightly modified form:
The reflectance and geometric masking shadowing are assumed to be constant for each photon that interacts with a microfacet on the surface, irrespective of the emitting light source, incident direction or microfacet orientation for a given viewing configuration.
We call this constant mean reflectance $\meanRefl$ and define it implicitly for the smooth BRDF model via
\begin{align}
	\label{eq:meanrefl}
	\Lo
	= \meanRefl \cdot \int_\domainHemi \ndfhol \cdot \Li \dx{\dirI}
	,
\end{align}
where $\meanRefl$, originally part of the integrand, is pulled out.
The mean reflectance $\meanRefl$ thus represents the ratio of reflected radiance divided by the integrated NDF weighted with the incoming radiance.
This quantity is preserved in the discrete model.
The discreteness affects only the number of microfacets, not their distribution or reflection properties, such that the reflectance and geometric masking shadowing terms are preserved.
\cref{eq:meanrefl} integrates the sum of all $K$ light sources indiscriminately.
%
Before proceeding to define the outgoing radiance for the discrete BRDF model,
we replace $\Li$ with its definition from \cref{eq:liksum}.
%
%
\begin{align}
	\Lo
	&= \meanRefl \cdot \sum_{k=1}^K \Lik \cdot \int_{\domainHemi} \ndfhol \cdot \wk \dx{\dirI}
	,
	\label{eq:smoothlopart}
\end{align}
where $\wk$ acts as indicator function for the domain of incoming light directions $\domainIk$ for each of the area lights.
\paragraph{Reflection Probability}
There is a distinct probability that any of the microfacets is oriented for a reflection from each of the area lights.
The probability for reflection from area light $k$ is
\begin{align}
	\probHk := \frac{\int_\domainHemi \ndfhol \cdot \wk \dx{\dirI}}{\int_\domainHemi \ndfhol \dx{\dirI}} = \frac{\intDk}{\intDHemi},
	\label{eq:probhk}
\end{align}
directly analogous to single area lights (\cref{eq:probh_def}).
\paragraph{Replacing the Discrete NDF}
%
%
The mean reflectance is assumed to be preserved when replacing the continuous NDF $\ndf$ with a corresponding discrete NDF $\ndfFoot$, i.e. $\Ex\ndfFoot = \ndf$ is required, which is fulfilled by the definition in \cref{sec:discretendf}.
%
We replace the integration of the continuous NDF $\ndf$ in \cref{eq:smoothlopart} with an integration of the discrete NDF $\ndfFoot$ to obtain the outgoing radiance for the glint BRDF:
\begin{align}
	\label{eq:iblglints:init}
	\LoFoot
	&= \meanRefl \cdot \sum_{k=1}^{K} \Lik \cdot \int_\domainHemi \ndfholFoot \cdot \wk \dx{\dirI} \\
	&\tikzboxed[thick,rounded corners=2pt]{
		= \Lo \cdot \frac{
		\sum_{k=1}^{K} \Lik \cdot M_k(\Nfoot, \probHvec)
	}{
		\Ex{\Nfoot} \cdot \sum_{k=1}^{K} \Lik \cdot \probHk
	} }\;,
	\label{eq:iblglints:fin}
\end{align}
where $\probHvec := \left(\probHk[1], \ldots, \probHk[K] \right)$ denotes the vector of reflection probabilities for all area lights, and $M_k(\Nfoot, \probHvec)$ is the sample count assigned to the $k$-th area light according to a multinomial distribution with $\Nfoot$ trials and success probabilities $\probHvec$.
A detailed derivation can be found in \cref{app:iblglintradiance}.
Each microfacet reflects light from exactly one of the area lights or a dummy light $k_\emptyset$ with $\Lik[k_\emptyset] = 0$ to account for shadowing and masking.
Therefore, we require that the individual sample counts actually refer to the different bins of a multinomial distribution, i.e. their sum is $\Nfoot$.
%
%




%
\section{Environment Map Filtering for Glint Rendering}
\label{sec:practice}
%


%
In \cref{sec:theory}, we have seen how to render glints under homogeneous area lights that partition the domain of incoming light directions.
This section describes how the environment map is split into multiple area lights, and how the probability for reflection is computed in practice.
We discretize the radiance from the environment map into a small finite set of four or eight brightness levels $\Lik$ as shown in \cref{fig:radiancelevels} (middle).
Each brightness level acts as an area light with constant radiance, but arbitrary shape.
For each area light, a scalar environment map indicates for each pixel whether it is part of the light source.
Applying environment map filtering \cite{karis2013real,lagarde2014moving} yields the corresponding integrated NDF $\intDk$, and by extension the respective probability of reflection $\probHk$, which is fed into the multinomial distribution in \cref{eq:iblglints:fin}.
The result is a simple and efficient approach based on real-time image-based lighting (\cref{sec:rtibl}).
%


%
\paragraph{Discrete Radiance Levels}
In contrast to the assumption from the previous section, environment maps are not homogeneous by nature.
Given the number of area lights $K$, we first determine the radiance levels $\Lik$ for $1 \leq k \leq K$.
We aim for a simple implementation to allow for dynamic environments and to limit the overhead during the environment map filtering step.
First, the minimum and maximum $\Li$ are determined.
A linear interpolation between the minimum and maximum in log-space leads to the definition
\begin{align}
	\Lik := \begin{cases}
		e^{\lerp\left(\log\min\Li, \log\max\Li, \alpha \right)} & \text{if $k > 1$} \\
		0 & \text{otherwise},
	\end{cases}
\end{align}
where $\alpha = \frac{k-1}{K-1}$.
Even for a white environment, microfacets might fail to reflect light because the visible reflecting microfacet area, which serves as an upper bound for the integrated NDF $\intD$, is less than the total microfacet area $\intDHemi$, leading to a strict inequality $\sum_{k=1}^{K} \probHk < 1$.
Therefore, the dimmest light source ($k=1$) serves the dual purpose of also representing invalid reflection directions below the horizon, for which we explicitly set $\Lik[1] = 0$.
For environments containing very dim or even black pixels, we found it useful to clip $\min \Li \geq \ell$, for a small user-defined $\ell \in [ 10^{-2}, 10^{-5} ]$, depending on the environment map.
This reduces the dynamic range of the illumination, preventing radiance levels from being wasted on over-representing dark regions that are unlikely to cause glints.
\paragraph{Fuzzy Indicator Functions}
The indicator functions $\wk$ are relaxed to store continuous values in the interval $[0,1]$ instead of discrete $\{0,1\}$, while preserving the partition of unity property $\sum_{k=1}^{K}\wk = 1$.
Thus, they are not actually indicator functions, but we call them fuzzy indicator functions instead.
This can be interpreted as a pixel being partially covered by all the light sources for which $\wk \neq 0$.
To determine $\wk$ for each pixel in the environment map, we find $j$ such that $\Lik[j] < \Li < \Lik[j+1]$.
The values of $\wk$ are then defined as
\begin{align}
	\wk := \begin{cases}
		1 - \frac{\Li - \Lik[j]}{\Lik[j+1] - \Lik[j]} & \text{if $k=j$,} \\
		\frac{\Li - \Lik[j]}{\Lik[j+1] - \Lik[j]} & \text{if $k=j+1$,} \\
		0 & \text{ otherwise.}
	\end{cases}
	\label{eq:wkdefpractice}
\end{align}
Effectively, $\wk$ is the linear interpolation factor such that $\Li = \lerp\left(\Lik[j], \Lik[j+1], \wk[j+1]\right)$ as illustrated in \cref{fig:radiancelevels} (bottom).
The computation of $\wk$ might also be performed in log-space.
However, our experiments show that a computation in linear space leads to fewer artifacts (c.f. \cref{fig:linlog}).
Notice that for increasing $K$ and given $\Li$, the interval spanned by the supporting radiance levels $\left[ \Lik[j], \Lik[j+1] \right]$ for which $\wk$ is non-zero shrinks and converges to the set $\left\{\Li\right\}$ for $K\rightarrow\infty$.

\iffigsinline

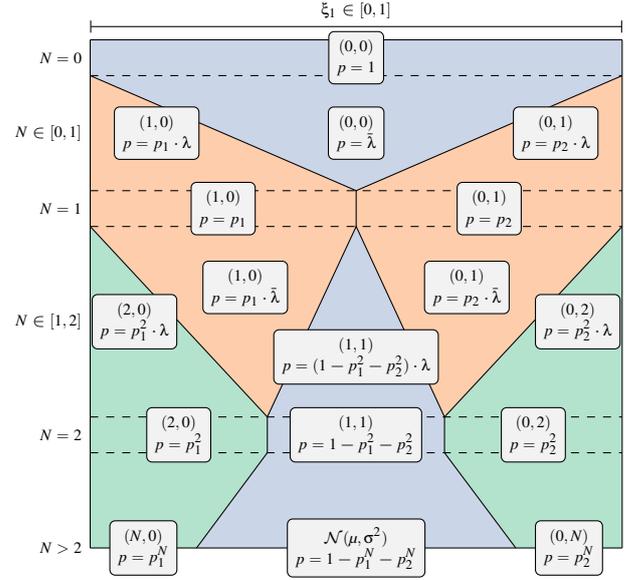
\begin{figure}
	\centering
	%
	%
	\tikzset{
		outerarea/.style={
			pastel2green,
		},
		innerarea/.style={
			pastel2red,
		},
		centerarea/.style={
			pastel2blue,
		},
		Nlabel/.style={
			font=\tiny,
		},
		Npnode/.style={
			midway,
			font=\tiny,
			align=center,
		},
		Npnodefrac/.style={
			Npnode,
			draw,
			line join=round,
			fill=black!5,
			rounded corners=2pt,
		},
		Npnodeint/.style={
			Npnode,
			draw,
			line join=round,
			fill=black!5,
			rounded corners=2pt,
		},
		up/.style={
			above=0.1cm,
		},
		down/.style={
			below=0.1cm,
		},
	}
	\def\plotwidth{7cm}
	\def\lineheight{0.75em}
	\newcommand{\Nfrac}{{\lambda}}
	\newcommand{\iNfrac}{{\bar{\lambda}}}
	\newcommand{\deflinecoord}[1]{
		\coordinate[above=\lineheight of #1] (#1T);
		\coordinate[below=\lineheight of #1] (#1B);
	}
	\begin{tikzpicture}		
		\coordinate (N0beg) at (0,0);
		\coordinate (N0end) at (\plotwidth,0);
		\deflinecoord{N0beg}
		\deflinecoord{N0end}

		\coordinate[below=2cm of N0beg] (N1beg);
		\coordinate[below=2cm of N0end] (N1end);
		\deflinecoord{N1beg}
		\deflinecoord{N1end}

		\coordinate[below=3cm of N1beg] (N2beg);
		\coordinate[below=3cm of N1end] (N2end);
		\deflinecoord{N2beg}
		\deflinecoord{N2end}

		\coordinate[below=1.5cm of N2beg] (NNbeg);
		\coordinate[below=1.5cm of N2end] (NNend);

		\coordinate (N1mid) at ($(N1beg)!0.5!(N1end)$);
		\deflinecoord{N1mid}
		\coordinate (N2left) at ($(N2beg)!0.333!(N2end)$);
		\coordinate (N2right) at ($(N2beg)!0.666!(N2end)$);
		\deflinecoord{N2left}
		\deflinecoord{N2right}
		\coordinate (NNleft) at ($(NNbeg)!0.2!(NNend)$);
		\coordinate (NNright) at ($(NNbeg)!0.8!(NNend)$);
		
		\coordinate (N01beg) at ($(N0beg)!0.5!(N1beg)$);
		\coordinate (N01left) at ($(N0beg)!0.5!(N1mid)$);
		\coordinate (N01right) at ($(N0end)!0.5!(N1mid)$);
		\coordinate (N01end) at ($(N0end)!0.5!(N1end)$);
		
		\coordinate (N12beg) at ($(N1beg)!0.5!(N2beg)$);
		\coordinate (N12left) at ($(N1beg)!0.5!(N2left)$);
		\coordinate (N12lmid) at ($(N1mid)!0.5!(N2left)$);
		\coordinate (N12rmid) at ($(N1mid)!0.5!(N2right)$);
		\coordinate (N12right) at ($(N1end)!0.5!(N2right)$);
		\coordinate (N12end) at ($(N1end)!0.5!(N2end)$);

		\fill[centerarea] (N0begT) -- (N0begB) -- (N1midT) -- (N0endB) -- (N0endT) -- cycle;
		\fill[innerarea] (N0begB) -- (N1midT) -- (N1midB) -- (N2leftT) -- (N1begB) -- (N1begT) -- cycle;
		\fill[innerarea] (N0endB) -- (N1midT) -- (N1midB) -- (N2right) -- (N1endB) -- (N1endT) -- cycle;
		\fill[centerarea] (N1midB) -- (N2leftT) -- (N2leftB) -- (NNleft) -- (NNright) -- (N2rightB) -- (N2rightT) -- cycle;
		\fill[outerarea] (N1begB) -- (NNbeg) -- (NNleft) -- (N2leftB) -- (N2leftT) -- cycle;
		\fill[outerarea] (N1endB) -- (NNend) -- (NNright) -- (N2rightB) -- (N2rightT) -- cycle;
		
		\node[Nlabel, left=0 of N0beg] {$N=0$};
		\node[Nlabel, left=0 of N01beg] {$N\in[0,1]$};
		\node[Nlabel, left=0 of N1beg] {$N=1$};
		\node[Nlabel, left=0 of N12beg] {$N\in[1,2]$};
		\node[Nlabel, left=0 of N2beg] {$N=2$};
		\node[Nlabel, left=0 of NNbeg] {$N>2$};
		
		
		\draw (N0begT) -- (NNbeg);
		\draw (N0endT) -- (NNend);
		
		\draw
			(N0begT) -- (N0endT)
			(NNbeg) -- (NNend);
		\draw[dashed]
			(N0begB) -- (N0endB)
			(N1begT) -- (N1endT)
			(N1begB) -- (N1endB)
			(N2begT) -- (N2endT)
			(N2begB) -- (N2endB);
		
		\draw (N0begB) -- (N1midT) -- (N0endB) (N1midT) -- (N1midB);
		\draw (N1begB) -- (N2leftT) -- (N1midB) -- (N2rightT) -- (N1endB);
		\draw (N2leftT) -- (N2leftB) -- (NNleft) (N2rightT) -- (N2rightB) -- (NNright);
		
		\path (N0beg) -- (N0end) node[Npnodeint] {$(0,0)$ \\ $p=1$};
		
		\path (N01beg) -- (N01left) node[Npnodefrac] {$(1,0)$ \\ $p=p_1 \cdot \Nfrac$};
		\path (N01left) -- (N01right) node[Npnodefrac] {$(0,0)$ \\ $p=\iNfrac$};
		\path (N01right) -- (N01end) node[Npnodefrac] {$(0,1)$ \\ $p=p_2 \cdot \Nfrac$};

		\path (N1beg) -- (N1mid) node[Npnodeint] {$(1,0)$ \\ $p=p_1$};
		\path (N1end) -- (N1mid) node[Npnodeint] {$(0,1)$ \\ $p=p_2$};
		
		\path (N12lmid) -- (N12rmid) node[Npnodefrac,down] {$(1,1)$ \\ $p=(1-p_1^2-p_2^2) \cdot \Nfrac$};
		\path (N12beg) -- (N12left) node[Npnodefrac] {$(2,0)$ \\ $p=p_1^2 \cdot \Nfrac$};
		\path (N12left) -- (N12lmid) node[Npnodefrac,up] {$(1,0)$ \\ $p=p_1 \cdot \iNfrac$};
		\path (N12rmid) -- (N12right) node[Npnodefrac,up] {$(0,1)$ \\ $p=p_2 \cdot \iNfrac$};
		\path (N12right) -- (N12end) node[Npnodefrac] {$(0,2)$ \\ $p=p_2^2 \cdot \Nfrac$};

		\path(N2beg) -- (N2left) node[Npnodeint] {$(2,0)$ \\ $p=p_1^2$};
		\path(N2left) -- (N2right) node[Npnodeint] {$(1,1)$ \\ $p=1-p_1^2-p_2^2$};
		\path(N2right) -- (N2end) node[Npnodeint] {$(0,2)$ \\ $p=p_2^2$};
		
		\path(NNbeg) -- (NNleft) node[Npnodefrac] {$(N,0)$ \\ $p=p_1^N$};
		\path(NNleft) -- (NNright) node[Npnodefrac] {$\mathcal{N}(\mu,\sigma^2)$ \\ $p=1-p_1^N-p_2^N$};
		\path(NNright) -- (NNend) node[Npnodefrac] {$(0,N)$ \\ $p=p_2^N$};
		
		\coordinate[above=5pt of N0begT] (xibeg);
		\coordinate[above=5pt of N0endT] (xiend);
		
		\draw[|-|] (xibeg)
		-- (xiend)
		node[above,midway,font=\tiny]{$\xi_1\in [0,1]$};
		
	\end{tikzpicture}
	\caption{
		Illustration of the dual-gating mechanism for an evaluation of $b(p, N)$.
		$p_1 := p$ is the probability for success, and $p_2 := 1-p$ is the probability for attaining a negative sample.
		For non-integer $N$, we use $\Nfrac:= N-\lfloor N\rfloor$ and $\iNfrac:= 1-\Nfrac$ for interpolation.
		Each box shows the outcome of the binomial sampling in the first row as the tuple $(\npos, \nneg)$ of positive and negative samples, and the probability for selecting that case given $N$.
		For $N\in\{0,1,2\}$ (dashed rows), the uniform random variable $\xi_1$ is used to select one of the discrete outcomes in accordance with the binomial distribution.
		For real-valued $N<2$, the probability for selecting one of the discrete outcomes of adjacent integer-valued $N$ is interpolated.
		%
		%
		The cases for $N=2$ are continuously extended for $N>2$, and a normal distribution is sampled in the case of sampling at least one positive and negative outcome.
	}
	\label{fig:gatingdiagram}
\end{figure}\fi
\iffigsinline

\begin{algorithm}
	\caption{Dual-gated Gaussian}
	\label{alg:dualgated}
	\makebox[1.2cm][l]{\textbf{Input:}} Number of trials $N \in \mathbb{R}_{\geq0}$, success probability $p$, \\
	\makebox[1.2cm][l]{} uniformly distributed random numbers $\xi_1,\xi_2 \in \mathcal{U}[0,1]$, \\
	\makebox[1.2cm][l]{\textbf{Output:}} Number of positive samples $\npos$ and negative samples $\nneg$.
	\begin{algorithmic}[1]
		\algnewcommand{\LineComment}[1]{\State \(\triangleright\) #1}
		\algnewcommand{\Continue}[0]{\State \textbf{continue}}
		\algnewcommand{\Assign}[3][]{\State \makebox[#1][l]{#2} $\gets$ #3 \vphantom{$P_{[]}^{N}$} }

		\Function{$b_\mathrm{dualgated}$}{$N$, $p$, $\xi_1$, $\xi_2$}
		
		\Assign[0.8cm]{$\Ngeqtwo$}{$\max(2,N)$}
		\Assign[0.8cm]{$c$}{$\max\left(1, \frac{10^{-3.54}}{p}\right)$}
		
		\vspace{0.3\baselineskip}
		\LineComment{Probabilities}
		\Assign[0.8cm]{$\Pallhit$}{$\clamp_{[0,1]}(N-1) \cdot p^\Ngeqtwo$}
		\Assign[0.8cm]{$\Pallhitneg$}{$\clamp_{[0,1]}(N-1) \cdot (1-c \cdot p)^{\frac{\Ngeqtwo}{c}}$}
		\Assign[0.8cm]{$\Ponehit$}{$\max(0, 1-|1-N|) \cdot p$}
		\Assign[0.8cm]{$\Ponehitneg$}{$\max(0, 1-|1-N|) \cdot (1-p)$}
		
		\vspace{0.3\baselineskip}
		\LineComment{Masks}
		\Assign[1.2cm]{$\maskallhit$}{$\chi( \xi_1 < \Pallhit )$}
		\Assign[1.2cm]{$\maskalloronehit$}{$\chi( \xi_1 < \Pallhit+\Ponehit )$}
		\Assign[1.2cm]{$\maskallhitneg$}{$\chi( 1 - \Pallhitneg \leq \xi_1 )$}
		\Assign[1.2cm]{$\maskalloronehitneg$}{$\chi( 1-\Pallhitneg-\Ponehitneg \leq \xi_1 )$}
		\Assign[0.4cm]{$\maskgauss$}{$\chi(N>1) \cdot \left(1-\maskalloronehit\right) \cdot \left(1-\maskalloronehitneg\right)$}
		
		\vspace{0.3\baselineskip}
		\LineComment{Gaussian}
		\Assign[0.4cm]{$\mu$}{$1 + (\Ngeqtwo-2) \cdot p$}
		\Assign[0.4cm]{$\sigma^2$}{$(\Ngeqtwo-2) \cdot p \cdot (1-p)$}
		\Assign[0.4cm]{$G$}{$\clamp_{[1,\Ngeqtwo-1)]}\left( \Phi^{-1}(\xi_2; \mu, \sigma^2) \right)$}
		\Assign[0.4cm]{$\bar{G}$}{$\Ngeqtwo - G$}
		
		\vspace{0.3\baselineskip}
		\LineComment{Final Result}
		\Assign[0.4cm]{$\npos$}{$\maskallhit \cdot (\Ngeqtwo-1) + \maskalloronehit + \maskgauss \cdot G$}
		\Assign[0.4cm]{$\nneg$}{$\maskallhitneg \cdot (\Ngeqtwo-1) + \maskalloronehitneg + \maskgauss \cdot \bar{G}$}
		
		\State \Return $(\npos, \nneg)$
		\EndFunction
	\end{algorithmic}
\end{algorithm}
\fi

\paragraph{Integrated NDF}
We treat $\wk$ as indicator functions for $\domainIk$ 
in the computation of the integrated NDF for each of the area lights.
\cref{fig:hemisphereplot} illustrates the shape of a smooth BRDF lobe, smooth NDF lobe and the filter kernel $\iblKernel$ defined in \cref{eq:rtibl} to filter the environment map in real-time image-based lighting \cite{karis2013real,lagarde2014moving}.
Since the filtering kernel is only a crude approximation of the BRDF lobe already, we reuse the same $\iblKernel$ to approximate the integrated NDF
\begin{align}
	\intDk
	&= \int_{\domainHk} \ndfhol \dx{\dirI}
	= \int_{\domainHemi} \ndfhol \cdot \wk \dx{\dirI} \\
	&\approx \ndfFGD \cdot \int_{\domainHemi} \iblKernel \cdot \wk \dx{\dirI},
	\label{eq:rtiblndf}
\end{align}
where $\ndfFGD := \int_\domainHemi \ndfhol \dx{\dirI}$ is the visible reflecting microfacet area,
equivalent to the directional albedo $\brdfFGD$ of the BRDF.
%
%
In practice, we effectively have $\ndfhol \sim \brdfnol$, both evaluated with respect to $\dirI$ given $\dirO$.
The only difference between \cref{eq:rtibl} and \cref{eq:rtiblndf} is the directional albedo $\brdfFGD$ for the reflected radiance, and the visible reflecting microfacet area $\ndfFGD$ for the integrated NDF.
Last but not least, the probability for a microfacet to reflect light from the k-th area light $\probHk$ is computed from the integrated NDF $\intDk$ as defined in \cref{eq:probhk}.
%



%
\section{Sampling Multinomial Distributions}
\label{sec:multinomial}
Computing the reflected radiance of the discrete BRDF from the environment map requires the sampling of a multinomial distribution in order to assign the individual microfacets inside of the pixel footprint to reflect light from exactly one of the area lights of assumed constant incoming radiance.
%
%
Sampling the multinomial distribution is implemented as a hierarchical sampling of binomial distributions.
In order to achieve constant runtime despite an arbitrarily large number of microfacets, sampling the binomial distribution is approximated by a gated Gaussian \cite{deliot2023real}.
However, such single-gated Gaussians are not well suited for approximating multinomial samplings, because the positive and negative outcomes are not treated symmetrically.
To address this, we introduce dual-gated Gaussians, where a gating decision is made both for having at least one positive sample, and also having at least one negative sample.
But first we address an issue with the existing gated Gaussian formulation that removes some artifacts with large area lights \cite{kneiphof2024real}.
This fix is a crucial prerequisite for our dual-gated Gaussian approximation.
Lastly, we also address a numerical instability for large $N$ and small $p$.
\paragraph{Single-Gated Gaussians}
Simply approximating a sampling from the binomial distribution with a Gaussian (clamped to $[0,N]$) does not yield satisfactory results, especially for low $p$.
The idea of gated Gaussians is to first test if there is at least one positive sample.
If not, the result immediately becomes $0$.
If there is at least one positive sample, we take it and distribute the remainder of the samples by a normal distribution with appropriate mean and variance.
Given two random numbers $\xi_1,\xi_2 \in \mathcal{U}[0,1]$, sampling the binomial distribution is thus approximated as
\begin{align}
	b(N,p) \approx \left\lfloor \chi\left(\xi_1 < \Panyhit\right) \cdot \clamp_{[1,N]}\left(\Phi^{-1}(\xi_2; \mu,\sigma^2)\right) \right\rfloor,
	\label{eq:oldgated}
\end{align}
where $\chi(\cdot)$ is the indicator function that yields $1$ if its argument is true, and $0$ otherwise.
Here, it represents the gating by sampling a Bernoulli trial with probability $\Panyhit = 1 - (1-p)^N$, the probability of sampling $b(N,p) \geq 1$.
$\Phi^{-1}(\xi_2; \mu, \sigma^2)$ is the inverse CDF of the normal distribution $\mathcal{N}(\mu, \sigma^2)$.
The mean $\mu = 1 + (N-1) \cdot p$ and variance $\sigma^2 = (N-1)\cdot p\cdot (1-p)$ are chosen to reflect one known positive sample, and $N-1$ not-yet determined samples.

\iffigsinline

\begin{figure}
	\centering
	\def\placeholdersize{0.15\linewidth}
	\def\plotsize{2.0cm}
	\newcommand\imgpath[1]{assets/smallp/#1.png}
	\pgfplotsset{
		mainplot/.style={
			clip=true,
			x dir=reverse,
			unbounded coords=jump,
			axis on top=true,
			axis lines=box,
			xmin=-16,xmax=0,
			ymin=0,ymax=16,
			width=\plotsize,height=\plotsize,
			scale only axis,
			xtick={-16,-12,-8,-4,0},
			ytick={0,4,8,12,16},
			tick label style={font=\scriptsize,inner sep=1,outer sep=1},
			tick style={major tick length=2},
			tick align=outside,
			xtick pos=left,
			ytick pos=left,
			xlabel=\small$\log_{10}p$,
			ylabel=\small$\log_{10}N$,
			every axis x label/.style={inner sep=1,outer sep=1,at={(ticklabel* cs:0.5,8)},anchor=north},
			every axis y label/.style={inner sep=1,outer sep=1,at={(ticklabel* cs:0.5,8)},anchor=south,rotate=90}
		},
		cmapplot/.style={
			clip=true,
			unbounded coords=jump,
			axis on top=true,
			axis lines=box,
			xmin=0.0,xmax=1.0,
			ymin=0,ymax=1,
			width=\plotsize,height=0.15cm,
			scale only axis,
			ytick=\empty,
			tick label style={font=\scriptsize,inner sep=1,outer sep=1},
			tick style={major tick length=2},
			tick align=outside,
			xtick pos=right
		},
	}%
	\subfloat[Target]{%
		\begin{tikzpicture}[node distance=.5em,every node/.style={inner sep=0,outer sep=0}]
			\node[at={(0,0)}] (plot) {};
			\begin{axis}[
				at={(plot)},
				anchor=north,
				mainplot,
				]
				\addplot graphics[xmin=-16,ymin=0,xmax=0,ymax=16,includegraphics={angle=90}] {\imgpath{val32ref}};
			\end{axis}
			\node[above=0.2em of plot,minimum height=16pt] (cmap) {};
			\begin{axis}[
				at={(cmap.south)},
				anchor=south,
				cmapplot,
				xtick={0,1},
				]
				\addplot graphics[ymin=0,xmin=0,ymax=1,xmax=1,includegraphics={angle=90}] {\imgpath{cmap_val}};
			\end{axis}
		\end{tikzpicture}%
		\label{fig:smallp:val}%
	}%
	\hspace{-4pt}%
	\subfloat[Error: FP32]{%
		\begin{tikzpicture}[node distance=.5em,every node/.style={inner sep=0,outer sep=0}, remember picture]
			\node[at={(0,0)}] (plot) {};
			\begin{axis}[
				at={(plot)},
				anchor=north,
				mainplot,
				]
				\addplot graphics[xmin=-16,ymin=0,xmax=0,ymax=16,includegraphics={angle=90}] {\imgpath{diff32}};
			\end{axis}
			\node[above=0.2em of plot,minimum height=16pt] (cmap) {};
			\begin{axis}[
				at={(cmap.south)},
				anchor=south,
				cmapplot,
				xtick={0,0.5,1},
				xticklabels={$0$,$10^{-4}$,$1$},
				]
				\addplot graphics[ymin=0,xmin=0,ymax=1,xmax=1,includegraphics={angle=90}] {\imgpath{cmap_diff}};
			\end{axis}
		\end{tikzpicture}%
		\label{fig:smallp:diff}%
	}%
	\hspace{-4pt}%
	\subfloat[Error: FP32 + Bias]{%
		\begin{tikzpicture}[node distance=.5em,every node/.style={inner sep=0,outer sep=0}]
			\node[at={(0,0)}] (plot) {};
			\begin{axis}[
				at={(plot)},
				anchor=north,
				mainplot,
				]
				\addplot graphics[xmin=-16,ymin=0,xmax=0,ymax=16,includegraphics={angle=90}] {\imgpath{diff32eps}};
			\end{axis}
			\node[above=0.2em of plot,minimum height=16pt] (cmap) {};
			\begin{axis}[
				at={(cmap.south)},
				anchor=south,
				cmapplot,
				xtick={0,0.5,1},
				xticklabels={$0$,$10^{-4}$,$1$},
				]
				\addplot graphics[ymin=0,xmin=0,ymax=1,xmax=1,includegraphics={angle=90}] {\imgpath{cmap_diff}};
			\end{axis}
		\end{tikzpicture}%
		\label{fig:smallp:diffeps}%
	}%
	\caption{
		Evaluation of $(1-p)^N$ using $\SI{32}{\bit}$ floating-point (FP32) arithmetic.
		(\subref{fig:smallp:val}) shows the true value of $(1-p)^N$.
		(\subref{fig:smallp:diff}) shows the error of a naive FP32 evaluation of $(1-p)^N$.
		(\subref{fig:smallp:diffeps}) shows the absolute difference between an FP32 evaluation of $(1-c \cdot p)^{\nicefrac{N}{c}}$ and the true value of $(1-p)^N$,
		where $c = \max\left(1, \nicefrac{\epsilon}{p}\right)$ and $\epsilon=10^{-3.54}$ is chosen optimally.
		Notice that despite evaluating a different quantity, the error to the original function is greatly reduced.
	}
	\label{fig:smallp}
\end{figure}
\fi

\paragraph{Gating Generalization}
The issue we found with previous implementations of gated Gaussians concerns scenarios where $N<1$.
As discussed earlier in \cref{sec:anisogrid}, the number of microfacets inside of the pixel footprint are distributed on a grid \cite{deliot2023real}.
%
%
Therefore, even for densely populated surfaces, in the vicinity of grid vertices, the number of microfacets that is distributed according to any of the neighboring vertices might attain values $N < 1$.
%
For sparsely populated surfaces, we naturally have $N < 1$.
Since the gated Gaussian is an approximation of the binomial distribution anyways, we generalize it to $N \in \mathbb{R}_{\geq0}$.
For $N>1$, it is sufficient to drop the floor operation from \cref{eq:oldgated}.
For $N<1$, the gating is modified in the following way:
We assume that there is a single microfacet on the surface, and $N$ describes the probability of that microfacet to reside inside of the pixel footprint.
For the binomial sampling this results in two independent binary decisions:
The first decision determines with probability $N$, if the microfacet is inside of the pixel footprint.
The second determines with probability $p$, if the microfacet oriented correctly.
For $N<1$, we simply replace the probability of the Bernoulli trial with $N \cdot p$, and make sure that the Gaussian part is clamped to exactly $1$.
The generalization to $N \in \mathbb{R}_{\geq0}$ for the gated Gaussian is thus
\begin{equation}
	\label{eq:gated}
	\begin{split}
		b(N,p) &\approx \chi\left(\xi_1 < \min(N \cdot p, \Panyhit) \right) \\
		&\quad \cdot \clamp_{[1,\Ngeqone]}\left(\Phi^{-1}(\xi_2; \mu,\sigma^2)\right),
	\end{split}
\end{equation}
where $\Ngeqone := \max(N, 1)$.
We prove that $\min(N \cdot p, \Panyhit)$ evaluates to $N\cdot p$ for $N<1$ and $\Panyhit$ otherwise in \cref{app:gatingprob}.
%


\iffigsinline\input{source/figure/result_sphere.tex}\fi

\iffigsinline\input{source/figure/result_whitefurnace.tex}\fi

\paragraph{Dual Gating}
Since the sampling of multinomial distributions is realized as a recursive sampling of binomial distributions, we are not just interested in the number of positive samples, but also symmetrically in the number of negative samples.
%
Unfortunately, the single-gating mechanism does not yield a symmetric distribution.
Therefore, we introduce the dual-gating mechanism, where an additional gating step is performed, such that \emph{both} cases of having at least one positive sample, and having at least one negative sample are considered explicitly.
In the following, we consider a binomial sampling $b(N, p)$ with $N \in \mathbb{R}_{\geq 0}$ trials and success probability $p$ for each trial.
The number of positive samples is denoted as $\npos$, and the number of negative samples as $\nneg$.
For $N > 2$, the dual-gating leads to three cases:
All samples are positive with probability $\Pallhit = p^N$, all samples are negative with probability $\Pallhitneg = (1-p)^N$, or a clamped Gaussian distribution is evaluated with remaining probability $\Pgauss = 1-\Pallhit-\Pallhitneg$.
%
%
The number of positive and negative samples are clamped to $[1,N-1]$, since we already know that there is at least one positive and negative sample from the preceding gating step.
For $N\in\{0,1,2\}$, the gating mechanism is used to sample the binomial distribution directly.
In the case of real-valued $N<2$, it is effectively dithered to the nearest integer before sampling the binomial distribution directly.
Thus, the positive and negative samples might not add up $\npos + \nneg \neq N$, making it necessary to track both $\npos$ and $\nneg$ explicitly.
%
%
The dithering is not necessary for $N > 2$, since the outcome is approximated by a normally distributed real value in $[1,N-1]$, which results in $\npos + \nneg = N$ by construction.
\cref{fig:gatingdiagram} illustrates the gating outcomes and their probability.
We identify 6 cases to consider during gating:
There are two cases with $\npos = 0$, where $\nneg \in \{1, \max(N, 2)\}$, and two complementary cases.
%
The two remaining cases yield either $\npos = \nneg = 0$ for $N<1$, or the clamped Gaussian for $N>1$.
\cref{alg:dualgated} details our implementation of the dual-gated Gaussian sampling that conveniently avoids any branching.
Note that only a single uniform random number is used for all gating decisions, plus one for sampling the normal distribution.
%


%
\paragraph{Numeric Gating Issues}
In our tests we noticed that there is a numeric instability when evaluating the gating.
It affects both our dual-gating mechanism, as well as the single-gating mechanism used for point-lights \cite{deliot2023real} and area lights \cite{kneiphof2024real}.
When evaluating the gating with small success probability $p$, the gating fails because $\Pallhitneg$ and $\Panyhit$ are computed incorrectly.
Due insufficient precision of $\SI{32}{\bit}$ floating-point values, $(1-p) \approx 1$ for small $p$.
Therefore, $(1-p)^N$ is calculated as $1^N = 1$ due to rounding errors, but $\lim_{N\rightarrow\infty} (1-p)^N = 0$.
We plot the error in \cref{fig:smallp}.
The incorrect result is obtained for $p<10^{-7.525}$ and $N>p^{-1}$.
Applying a small bias to the computation, the error becomes negligible.
%

%
First, the exponent of $(1-p)^N$ is expanded with $\frac{c}{c}$, where $c$ is to be determined.
Next, the Taylor series expansion of $(1-p)^c$ is taken at $p=0$ as follows:
\begin{align}
	(1-p)^N = \left[(1-p)^c\right]^{\frac{N}{c}} = \left[ 1 - c \cdot p + \mathcal{O}\left(p^2\right) \right]^{\frac{N}{c}}.
\end{align}
Since $p$ is small, we drop the $\mathcal{O}\left(p^2\right)$ terms and get a formula similar to the original $(1-p)^N$.
We effectively introduce a new scaling factor $c$, acting upon $p$ and $N$:
\begin{align}
	(1-p)^N \approx (1 - c \cdot p)^{\frac{N}{c}}.
\end{align}
Since our goal is to work around small $p$, we certainly want $c \geq 1$.
Let $\epsilon>0$ be the smallest value for which the computation $(1-\epsilon)^N$ is sufficiently accurate.
We choose $c := \max(1, \nicefrac{\epsilon}{p})$.
This results in $(1-\epsilon)^{\nicefrac{N\cdot p}{\epsilon}}$ for $p<\epsilon$, and an unmodified $(1-p)^N$ for sufficiently large $p>\epsilon$.
Experimentally, we found $\epsilon = 10^{-3.54}$ to be optimal for FP32 arithmetic and $(N, p) \in [1, 10^{16}] \times [10^{-16}, 1]$.
%




%
\section{Evaluation}
\label{sec:evaluation}
We implement our method in the High Definition Render Pipeline 14.0.10 for Unity3D 2022.3 based on previous work \cite{kneiphof2024real,deliot2023real}.
Our implementation is restricted to isotropic GGX microfacet BRDFs with the Smith shadowing assumption \cite{heitz2014understanding}, and Schick's approximation for the Fresnel term \cite{schlick1994inexpensive}.
We only consider direct illumination, and occlusion is not taken into account.
Also, we do not consider any multiple scattering reflections in our implementation.
To generate reference renderings, we implemented explicitly distributed microfacets (similar to \cite{jakob2014discrete}) in Mitsuba 3 \cite{Mitsuba3}.

\iffigsinline\input{source/figure/result_wang2020.tex}\fi

\paragraph{Teaser Image}
In \cref{fig:teaser}, we show some example scenes containing glittery appearance.
The woman wears a dress, exhibiting a sparkling effect sparsely distributed on the surface.
On the car, we model the glints due to metallic flakes embedded in the paint, and last but not least some glittering ornamental balls.
All materials are modeled as a mixture of our glint model with some diffuse and smooth specular components.

\iffigsinline\input{source/figure/precision.tex}\fi

\iffigsinline\input{source/figure/linlog.tex}\fi

\paragraph{Comparison with Ground Truth}
In order to demonstrate the robustness of our method, we compare against ground-truth renderings, where microfacets are distributed explicitly over the surface (c.f. \cref{fig:result_sphere}).
We show low to high microfacet densities, low to high surface roughness and three different environments.
Our approximation is overall very similar to the reference.
Only for $K=4$ radiance levels, a difference in our method becomes apparent.
This is usually the case when the dynamic range of the environment map is too high, and the chosen radiance levels do not represent the true radiance distribution well enough.
The supplemental video contains an animated version of \cref{fig:result_sphere}, as well as an animated swiveling piece of cloth.
The videos show that the angular lifetime of the reference glints depends on the environment map.
Glints from low-frequent illumination posses a longer lifetime than those from high-frequent illumination.
Since our method does not take this into account,
the angular lifetime of our glints deviates from ground truth.
\paragraph{White Furnace}
We validate the behaviour of our glint model in the white furnace environment against the smooth BRDF model \cref{fig:result_whitefurnace}.
Some percentage of the microfacets are not oriented for any reflection from the environment map, i.e. their reflection direction points below the horizon.
In reality, multiple scattering would occur \cite{chermain2019glint}, but we only consider single scattering in our real-time approximation.
When the microfacet density is low, fewer but brighter glints are produced, since the incident radiance is distributed over fewer microfacets.
Averaging multiple realizations using the same microfacet density still converges to the appearance of the smooth model.

\iffigsinline\input{source/figure/result_rgb.tex}\fi

\paragraph{Comparison with Previous Work}
\cref{fig:result_wang2020} compares our method to the idea presented by Wang et al. \cite{wang2020real} in previous work.
We took their idea and implemented it on top of the anisotropic grid \cite{deliot2023real}, which avoids the traversal of a spatial quad tree, and uses gated Gaussians instead of explicit sums for sampling binomial distributions.
This makes it possible to always use the micro-scale model (c.f. Eq. 20 in \cite{wang2020real}).
Similar to our method, they modulate the reflected radiance of a smooth surface.
However, the probability for reflection of a single microfacet only depends on a user-adjustable parameter $\gamma$, and is independent of the environment map contents, resulting in a globally uniform glint appearance.
In contrast, the glints produced by our method depend on the shape of the light sources captured in the environment map, automatically producing stronger glints for smaller, brighter regions.
\paragraph{Texture Precision}
We compare different texture formats for the additional $\wk$ stored in the prefiltered environment map in \cref{fig:precision}.
The $\SI{8}{\bit}$ formats simply have insufficient precision.
The $\SI{16}{\bit}$ formats and $\SI{32}{\bit}$ SFloat format produce very similar results, such that the higher storage costs of the $\SI{32}{\bit}$ format are not justified.
Depending on the scenario, we found the $\SI{16}{\bit}$ UNorm format to be visually marginally superior to the $\SI{16}{\bit}$ SFloat format.
\paragraph{Linear vs Logarithmic Radiance}
We determine the brightness $\Lik$ of the area lights in log space.
However, the assignment $\wk$ of each pixel in the environment map is performed in linear space.
\cref{fig:linlog} compares weighting computation in linear vs. log space.
The difference is subtle, but for $K=4$ levels, using the log space results in a slightly dark halo around the reflection.
For increasing $K$ however, the choice does not matter because the distances over which to interpolate vanish.
\paragraph{Colored Illumination}
The primary limitation of our method stems from the assumption that the environment map only varies in brightness, not chromaticity.
We illustrate this by placing three colored light sources at different distance to each other in \cref{fig:result_rgb}.
When the light sources are close together, we are not able to differentiate which of the lights should be reflected, and the glints become white.
When the light sources and their specular reflection is clearly separated, our glints produce the correct color again.
\paragraph{Dual-Gated Gaussians}
We validate our dual-gated Gaussian distribution against a single-gated Gaussian distribution and a binomial distribution $b(N,p)$ in \cref{fig:gatingplot}.
For large $N$ all three distributions are similar.
However, for small $N$, the single-gated Gaussian appears different from the dual-gated Gaussian which produces a distribution close to the reference.

\iffigsinline

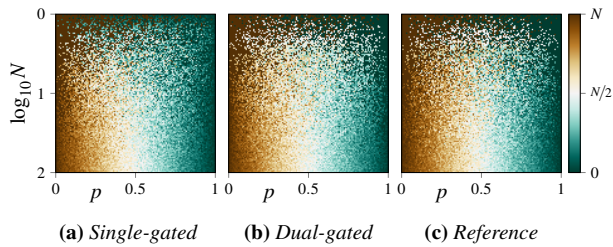
\begin{figure}
	\centering
	\def\placeholdersize{0.15\linewidth}
	\def\plotsize{2.1cm}
	\newcommand\imgpath[1]{assets/gatingplot/#1.png}
	%
	%
	\pgfplotsset{
		mainax/.style={
			anchor=center,
			clip=true,
			y dir=reverse,
			unbounded coords=jump,
			axis on top=true,
			axis lines=box,
			xmin=0,xmax=1,
			ymin=0,ymax=2,
			width=\plotsize,height=\plotsize,
			scale only axis,
			xtick={0,0.5,1},
			tick label style={font=\scriptsize,inner sep=1,outer sep=1},
			tick style={major tick length=2},
			tick align=outside,
			xtick pos=left,
			ytick pos=left,
			xlabel=\small$p$,
			every axis x label/.style={inner sep=1,outer sep=1,at={(ticklabel* cs:0.25,4)},anchor=north},
			every axis y label/.style={inner sep=1,outer sep=1,at={(ticklabel* cs:0.5,8)},anchor=south,rotate=90},
		},
		singleax/.style={
			mainax,
			ytick={0,1,2},
		},
		dualax/.style={
			mainax,
			ytick=\empty,
		},
		refax/.style={
			mainax,
			ytick=\empty,
		},
		cmapax/.style={
			clip=true,
			unbounded coords=jump,
			axis on top=true,
			axis lines=box,
			xmin=0,xmax=1,
			ymin=0,ymax=1,
			width=0.15cm,height=\plotsize,
			scale only axis,
			xtick=\empty,
			tick label style={font=\scriptsize,inner sep=1,outer sep=1},
			tick style={major tick length=2},
			tick align=outside,
			ytick pos=right,
			ytick={0,0.5,1},
			yticklabels={$0$,$\nicefrac{N}{2}$,$N$},
		},
		plot graphics/mainimg/.style={
			xmin=-0,ymin=0,xmax=1,ymax=2,
		},
	}%
	\tikzset{
		mainaxnode/.style={
			at={(0,0)},
			minimum size=\plotsize,
		},
		cmapaxnode/.style={
			right=0.2em of #1,
			minimum width=0.15cm,
			minimum height=\plotsize,
		},
		Nlabelnode/.style={
			left=6pt of #1,
			rotate=90,
			anchor=south,
			font=\small
		},
		plabelnode/.style={
			below left= 2pt and -16pt of #1,
			font=\small
		},
	}
	\subfloat[Single-gated]{%
		\begin{tikzpicture}[node distance=.5em,every node/.style={inner sep=0,outer sep=0}, remember picture]
			\node[mainaxnode] (singleaxnode) {};
			\begin{axis}[
				at={(singleaxnode)},
				singleax,
				]
				\addplot graphics[mainimg] {\imgpath{gated}};
			\end{axis}
		\end{tikzpicture}%
		\label{fig:gating:single}%
	}%
	\subfloat[Dual-gated]{%
		\begin{tikzpicture}[node distance=.5em,every node/.style={inner sep=0,outer sep=0}, remember picture]
			\node[mainaxnode] (dualaxnode) {};
			\begin{axis}[
				at={(dualaxnode)},
				dualax,
				]
				\addplot graphics[mainimg] {\imgpath{dualgated}};
			\end{axis}
		\end{tikzpicture}%
		\label{fig:gating:dual}%
	}%
	\subfloat[Reference]{%
		\begin{tikzpicture}[node distance=.5em,every node/.style={inner sep=0,outer sep=0}, remember picture]
			\node[mainaxnode] (refaxnode) {};
			\begin{axis}[
				at={(refaxnode)},
				refax,
				]
				\addplot graphics[mainimg] {\imgpath{ref}};
			\end{axis}
		\end{tikzpicture}%
		\label{fig:gating:ref}%
	}%
	\begin{tikzpicture}[remember picture,overlay]
		\node [Nlabelnode=singleaxnode] {$\log_{10}N$};
		\node[cmapaxnode=refaxnode] (cmapnode) {};
		\begin{axis}[
			at={(cmapnode)},
			anchor=center,
			cmapax,
			]
			\addplot graphics[ymin=0,xmin=0,ymax=1,xmax=1] {\imgpath{cmap}};
		\end{axis}
	\end{tikzpicture}
	\caption{
		Comparison of the single-gated and dual-gated Gaussian approximation of the binomial distribution $b(N, p)$.
		Differences are most apparent for low $N$.
		The single-gated approximation is asymmetric by construction.
		The dual-gated approximation is symmetric, which is beneficial when used in a multinomial distribution. Overall, it is much closer to the reference due to the additional gating step.
	}
	\label{fig:gatingplot}
\end{figure}
\fi


\begin{table}
	\centering
	\newcommand{\tablepath}[2]{assets/runtime/#1/table_flat_#2.csv}
	\setlength\tabcolsep{3.5pt}
	%
	%
	\begin{tabular}{lcccc}
		\toprule
		Light & Smooth & \citeq{wang2020real}* & $K=4$ & $K=8$ \\


		\midrule
		&\multicolumn{4}{l}{\tiny RTX 4090 @ $3840\times2160$ \hfill [\unit{\milli\second}]} \\
		\midrule
		\csvreader[head to column names,late after line=\\, late after last line=\\]{\tablepath{hochstaufen_gc}{ibl}}{}
			{ \light & $\smoothE \pm \smoothD$
				& $\wangUniformUshortE \pm \wangUniformUshortD$
				& $\glintIVuniformUshortE \pm \glintIVuniformUshortD$
				& $\glintVIIIuniformUshortE \pm \glintVIIIuniformUshortD$
			}
		\csvreader[head to column names,late after line=\\, late after last line=\\]{\tablepath{hochstaufen_gc}{dir}}{}
			{ \light & $\smoothE \pm \smoothD$
				& {\tiny \phantom{$\leftarrow$} same $\rightarrow$}
				& $\glintUniformE \pm \glintUniformD$
				& {\tiny $\leftarrow$ same \phantom{$\rightarrow$}}
			}

		\midrule
		&\multicolumn{4}{l}{\tiny RTX 4070M @ $3840\times2160$ \hfill [\unit{\milli\second}]} \\
		\midrule
		\csvreader[head to column names,late after line=\\, late after last line=\\]{\tablepath{xpswin11_gc}{ibl}}{}
			{ \light & $\smoothE \pm \smoothD$ %
				& $\wangUniformUshortE \pm \wangUniformUshortD$ %
				& $\glintIVuniformUshortE \pm \glintIVuniformUshortD$ %
				& $\glintVIIIuniformUshortE \pm \glintVIIIuniformUshortD$ %
			}
		\csvreader[head to column names,late after line=\\, late after last line=\\]{\tablepath{xpswin11_gc}{dir}}{}
			{ \light & $\smoothE \pm \smoothD$
				& {\tiny \phantom{$\leftarrow$} same $\rightarrow$}
				& $\glintUniformE \pm \glintUniformD$
				& {\tiny $\leftarrow$ same \phantom{$\rightarrow$}}
			}

		\bottomrule
	\end{tabular}
	\caption{
		We measure the runtime when rendering a full screen plane illuminated by an environment map only and a directional light for comparison.
		The average runtime per frame and standard deviation is shown for two GPUs.
		%
		%
		We distinguish the static environment map scenario, and the dynamic case where the environment map is prefiltered in every frame.
		The columns show the continuous BRDF without glints (Smooth),
		the previous real-time image-based lighting method \cite{wang2020real}, which we reimplement based on \citeq{deliot2023real} for increased performance over the original,
		as well as our method using $K \in \{4,8\}$ radiance levels.
		The implementation for directional lights is identical to \citeq{deliot2023real} for all three glint variants.
	}
	\label{tab:runtime}
\end{table}

\paragraph{Improved Single-Gated Gaussians}
In preparation for our dual-gated Gaussian formulation, we propose an improvement to the single-gated Gaussians.
In previous work \cite{deliot2023real,kneiphof2024real}, $\Nfoot \in \mathbb{N}$ is required theoretically, but in practice $\Nfoot < 1$ occurs frequently.
This discrepancy leads to visual artifacts when $\probH$ is large, which we construct by placing a large area light above a surface with low roughness \cref{fig:result_nlt1}.
For high $\Nfoot$, the appearance still converges to the smooth BRDF.
However, for lower $\Nfoot$, artifacts become visible, and the brightness of the glints is overestimated.
With our modification to handle $\Nfoot < 1$, those artifacts are alleviated.

\iffigsinline\input{source/figure/result_nlt1.tex}\fi

\paragraph{Numeric Gating Issues}
In \cref{fig:result_smallp} we show an example of artifacts that are caused by errors to the floating-point arithmetic in the $(1-\probH)^\Nfoot$ computation.
We construct a scenario with very small $\probH$ using a narrow point light viewed in an off-specular configuration on a surface with low roughness.
The banding artifacts do not disappear when increasing $\Nfoot$, such that the appearance should converge towards that of the equivalent smooth BRDF.
With our simple modification, the appearance converges for increasing $\Nfoot$, and the banding artifacts are avoided.
\paragraph{Runtime}
A comparison of runtime performance is shown in \cref{tab:runtime}.
The timings were recorded using an Nvidia RTX 4090 GPU and a mobile Nvidia RTX 4070 GPU power-limited to $\SI{35}{\watt}$, running Windows 11 with Unity's DirectX 12 backend.
Our glint-rendering technique is compared to the shading of a smooth surface, our implementation of previous work \cite{wang2020real}, and a illumination from a single directional light source \cite{deliot2023real}.
All listed image-based lighting techniques have some overhead compared to simply rendering a directional light source.
Our implementation of the previous technique \cite{wang2020real} has some negligible overhead over rendering only the directional light source, despite being functionally equivalent.
Our method with $K=4$ is slightly slower because of the additional environment map data and the more complex multinomial sampling.
For $K=8$, the runtime scales superlinearly, which is not fully explained by the increased complexity of sampling the multinomial distribution.
%
Profiling our implementation reveals that \SI{55}{\percent} of GPU stalls are due to instruction cache misses for $K=8$ (None for $K=4$).
This suggests that reducing the size of the shader program, e.g. by disabling code inlining in the right places, could improve the performance.
Filtering the environment map in each frame increases the runtime slightly over a static environment map, but not by much.

\iffigsinline\input{source/figure/result_smallp.tex}\fi


%
\section{Conclusion}
\label{sec:conclusion}
In this paper we have presented a novel approach for rendering glints under image-based lighting in real time.
It produces renderings that are close to reference with little runtime overhead over rendering glints from a single directional light.
The implementation of our method nicely integrates with the previous work in the same theoretical framework \cite{deliot2023real,kneiphof2024real}.
Additionally, we have resolved two technical limitations that also affect glints under directional and area light sources.
\paragraph{Assumptions}
The core of our idea is that the environment map is well approximated by few homogeneous regions, and chromatic variations in the illumination are negligible in practice.
As in previous glint-rendering techniques, the reflection properties of glints are assumed to be constant, irrespective of their orientation.
For image-based lighting this is quite a stretch, but we demonstrated reasonable appearance.
We assume both the NDF and the BRDF lobe to be approximated by the filter kernel used in real-time image-based lighting, and therefore to be identical in shape.
\paragraph{Limitations \& Future Work}
One aspect that has been overlooked so far is the interaction of glint BRDFs with diffuse and smooth BRDF lobes.
In \cref{fig:teaser}, a simple linear blending between the three BRDF lobes is used.
We suppose that a more rigorous solution would be to embed the concept of glints into layered BRDF frameworks \cite{belcour2018efficient}.
In this paper, we only show results for isotropic microfacet distributions.
However, by drawing multiple samples from the prefiltered environment map \cite{cocco2024anisotropic}, anisotropic distributions should be easily accommodated.
We only consider unshadowed illumination from the environment map.
Multiplying occlusion estimates\cite{heitz2018combining} directly to the \emph{output} of our method would lead to incorrect results.
For correct occlusion, the probabilities used to sample the multinomial distributions should be attenuated, depending on which regions of the environment map are occluded.
Adding support for chromatic variations of individual glints would be an interesting avenue for future work.
Both regarding the reflection properties of individual glints, as well as chromatic variations in the environment map.
We presented a very minimal approach for approximating the environment map by homogeneous regions.
More elaborate clustering approaches could improve the behaviour of our method for complex environments, including support for color-valued radiance.
Our method does not take into account the angular lifetime of glints.
Depending on the roughness and environment map, glints might be stable or rapidly pop in and out of existence.
For directional and area light sources, the lifetime can be controlled manually by adjusting the grid resolution with respect to the halfway vector \cite{deliot2023real}.
With real-time image-based lighting however, we do not have access to a meaningful halfway vector.
Controlling the angular lifetime of glints in a principled way is an interesting direction for future research.
%


%
\section*{Acknowledgements}
The authors sincerely thank
\ifanonymous
\enquote{redacted due to anonymous submission}
\else
Laurent Belcour, Xavier Chermain, and Thomas Deliot
\fi
for their valuable discussions and insightful feedback.
The models shown in \cref{fig:teaser} are provided under CC-BY license from
the \textsc{Walking Dress} (\url{https://skfb.ly/ptF8A}) by User \enquote{gulyayeva},
the \textsc{Sports Car} (\url{https://github.com/MirageYM/3DModels}) by Yasutoshi Mori,
and the \textsc{Christmas Ornaments} (\url{https://skfb.ly/oo7ro}) by Naira Arakelyan.
%

\bibliographystyle{eg-alpha-doi}
\bibliography{IBLGlints-bibliography}


\appendix


%
\section{Outgoing Radiance of Glint Model as Multinomial Sampling}
\label{app:iblglintradiance}
This appendix details the transformation steps from \cref{eq:iblglints:init} to \cref{eq:iblglints:fin} for the outgoing radiance of a BRDF with discrete NDF, where the incoming light is partitioned into regions of constant radiance according to \cref{eq:liksum}.
%
The mean reflectance $\meanRefl$ implicitly defined in \cref{eq:meanrefl} can be simplified as
\begin{align}
	\meanRefl &= \frac{\int_\domainH \brdfnol \cdot \Li \dx{\dirI}}{\int_\domainH \ndfhol \cdot \Li \dx{\dirI}} \\
	&= \frac{\Lo}{ \int_\domainH \ndfhol \cdot \sum_{k=1}^{K} \wk \cdot \Lik \dx{\dirI} } \\
	&= \frac{\Lo}{ \sum_{k=1}^{K} \Lik \cdot \int_\domainH \ndfhol \cdot \wk \dx{\dirI} } \\
	&= \frac{\Lo}{ \sum_{k=1}^{K} \Lik \cdot \intDk }.
	\label{eq:meanrefl:simple}
\end{align}
%
%
Plugging the simplified mean reflectance (\cref{eq:meanrefl:simple}) and the binomial version of the integral over the discrete NDF \cref{eq:binomial_intd} into the outgoing radiance of the glint model (\cref{eq:iblglints:init}), we get
\begin{align}
	\LoFoot
	&= \meanRefl \cdot \sum_{k=1}^{K} \Lik \cdot \int_\domainHemi \ndfholFoot \cdot \wk \dx{\dirI} \\
	&= \frac{\Lo}{\sum_{k=1}^{K} \Lik \cdot \intDk }
	\cdot \sum_{k=1}^{K} \Lik \cdot \frac{\intDHemi \cdot M_k(\Nfoot, \probHvec)}{\Ex{\Nfoot}}.
\end{align}
Since we are dealing with a sum of discrete NDF integrals over disjoint domains $\domainHk$, we know that each of the $\Nfoot$ trials will result in a positive outcome for only one of the $K$ light sources.
Therefore, we replace the binomial distributions $b\left(\Nfoot, \probHk\right)$ with a single multinomial distribution, where the resulting sample count for the $k$-th light is denoted as $M_k(\Nfoot, \probHvec)$ with the vector of probabilities $\probHvec := \left(\probHk[1], \ldots, \probHk[K]\right)$ for each bin.
Further transformation yields
\begin{align}
	\LoFoot &= \Lo \cdot \frac{
		\sum_{k=1}^{K} \Lik \cdot M_k(\Nfoot, \probHvec)
	}{
		\Ex{\Nfoot} \cdot \sum_{k=1}^{K} \Lik \cdot \frac{\intDk}{\intDHemi}
	} \\
	&= \Lo \cdot \frac{
		\sum_{k=1}^{K} \Lik \cdot M_k(\Nfoot, \probHvec)
	}{
		\Ex{\Nfoot} \cdot \sum_{k=1}^{K} \Lik \cdot \probHk
	},
\end{align}
which is equal to \cref{eq:iblglints:fin}.
The multinomial sampling is used in a weighted sum with the corresponding radiance values $\Lik$, divided by the expected value of the weighted sum.
This quantity then simply modulates the outgoing radiance of the equivalent smooth BRDF model.
%


%
\section{Single-Gating Probability}
\label{app:gatingprob}
We propose to replace the probability for the Bernoulli trial in single gated Gaussians in \cref{eq:gated} to support $N<1$ when sampling binomial distributions $b(N,p)$.
This appendix proves that $\min(N \cdot p, \Panyhit)$ evaluates to $N\cdot p$ for $N<1$ and $\Panyhit$ otherwise.

For $N\in\mathbb{R}_{\geq0}$ and $p\in[0,1]$, consider the following difference and its derivative with respect to $p$
\begin{align}
	f(N,p) &= \overbrace{1 - (1-p)^N}^\Panyhit - N \cdot p, \\
	\derivative{f(N,p)}{p} &= N \cdot \left( (1-p)^{N-1} - 1 \right)
\end{align}
Notice that $f(N,p)=0$, i.e. $N \cdot p = \Panyhit$, if $N \in \{0, 1\}$ or $p=0$.
\begin{itemize}
\item 
	For $N > 1$, we observe $\derivative{f(N,p)}{p} < 0$.
	%
	With $f(N,0) = 0$, we get $f(N,p) < 0$, i.e. $\min(N \cdot p, \Panyhit) = \Panyhit$.

\item
	For $N \in (0,1)$, we observe $\derivative{f(N,p)}{p} > 0$.
	%
	With $f(N,0) = 0$, we get $f(N,p) > 0$, i.e. $\min(N \cdot p, \Panyhit) = N\cdot p$.
\end{itemize}



\end{document}